\documentclass[11pt,preprint]{aastex}

\newcommand{\lam}{$\lambda$}

\newcommand{\kms}{km~s$^{-1}$}
\newcommand{\as}{${^\prime}{^\prime}$}

\renewcommand{\ion}[2]{#1\,{\sc #2}}

\def\ionx[#1 #2]{#1\,{\sc #2}}

\shorttitle{}
\shortauthors{}

\begin{document}


\title{The 2014 March 29 X-flare: sub-arcsecond resolution observations of Fe XXI \lam1354.1}


\author{Peter R. Young}
\affil{College of Science, George Mason University, Fairfax, VA 22030,
USA}
\author{Hui Tian}
\affil{Harvard-Smithsonian Center for Astrophysics, 60 Garden Street,
  Cambridge, MA 02138, USA}
\author{Sarah Jaeggli}
\affil{Department of Physics, Montana State University, 
  P.O. Box 173840, Bozeman, MT 59717, USA}



\begin{abstract}
The Interface Region Imaging Spectrometer
(IRIS) is the first solar instrument to observe $\sim 10$~MK
plasma at subarcsecond spatial resolution through imaging spectroscopy
of the \ion{Fe}{xxi} \lam1354.1 forbidden line.
IRIS observations of the X1 class flare that occurred on 2014
March 29 at 17:48~UT  reveal \ion{Fe}{xxi}
emission from both the flare ribbons and the post-flare loop arcade.
\ion{Fe}{xxi} appears at all of the chromospheric ribbon
  sites, although typically with a delay of one raster (75~seconds)
  and sometimes offset by up to 1\as. 100--200~\kms\ blue-shifts are
  found  at the brightest ribbons, suggesting hot plasma upflow into the corona.
The
\ion{Fe}{xxi} ribbon emission is compact with a spatial
extent of $< 2$\as, and can extend beyond the chromospheric ribbon
locations. Examples are found of both decreasing and increasing
blue-shift in the direction away from the ribbon locations, and 
blue-shifts were present for at least 6~minutes after the
flare peak. The post-flare loop arcade, seen in  Atmospheric Imaging Assembly (AIA) 131~\AA\ 
filtergram images that are dominated by \ion{Fe}{xxi}, exhibited
bright loop-tops with an asymmetric
intensity distribution. The sizes of the loop-tops are resolved by
IRIS at $\ge 1$\as, and line widths in the loop-tops are not
broader than in the loop-legs
suggesting the  loop-tops are not sites of enhanced
turbulence. Line-of-sight speeds in the loop arcade are typically
$<10$~\kms, and mean non-thermal  motions fall from 43~\kms\ at the
flare peak to 26~\kms\ six minutes later. 
If the average velocity in the loop arcade is
assumed to be at rest, then it implies a new reference wavelength for the
\ion{Fe}{xxi} line of $1354.106\pm 0.023$~\AA.
\end{abstract}





\section{Introduction}\label{sect.intro}

The $^3P_0$--$^3P_1$ ground transition of \ion{Fe}{xxi} gives rise to
an emission line at 1354.1~\AA\ that 
was first identified from \emph{Skylab} S082B spectra
by \cite{doschek75}. It is the strongest emission line formed at
temperatures $\ge 10$~MK that is found longward of the hydrogen Lyman limit,
and it is of great interest for studying plasma dynamics during flares
on account of the much higher spectral resolution possible at
ultraviolet wavelengths compared to X-ray wavelengths.
The first observations of \cite{doschek75} demonstrated for one flare that the line width
decreased as the flare evolved, and Doppler motions did not exceed
20~\kms. \citet{cheng79} studied 17 \emph{Skylab} flares and used \lam1354.1 to
determine non-thermal velocities ranging between 0 and 60~\kms. Note
that the S082B instrument had a 2\as\ $\times$ 60\as\ slit but no spatial
resolution along the slit, thus the spatial resolution of the
instrument depended on the spatial extent of the observed flare.

\citet{mason86} analyzed seven flares observed with the
Ultraviolet Spectrometer and Polarimeter (UVSP) on board the Solar
Maximum Mission (SMM), and found examples of asymmetries in the line
profile during the rise phase of flares that indicated plasma
upflowing at speeds of up to 200~\kms. Smaller asymmetries were also
found during the soft X-ray peaks of the flares, suggesting
evaporation was continuing at this stage of the flare. The spatial
resolution of UVSP was determined by the size of the entrance slit
used, which could be as small as 3\as\ $\times$ 3\as.

The Solar Ultraviolet Measurement of Emitted Radiation (SUMER)
instrument on board the Solar and Heliospheric Observatory (SOHO)
observed the \ion{Fe}{xxi} line, but only in off-limb 
flares. A M7.6 flare on 1999 May 9 was captured with a full spectrum
scan by SUMER and studied by \citet{feldman00},
\citet{innes01} and \citet{landi03}. The \ion{Fe}{xxi} line was
observed about 3~hours after the flare peak, and \citet{feldman00}
derived a new rest wavelength of $1354.064\pm 0.020$~\AA\ for the
line, and \citet{landi03} found non-thermal broadening 
corresponding to 30--40~\kms.  \ion{Fe}{xxi} \lam1354.1 was also recorded in the
X-class flare of 2002 April 21, which produced a supra-arcade of hot
plasma that displayed dynamic, dark voids
\citep{innes03b,innes03a}. Blueshifts of up to 1000~\kms\ were seen in
this data-set, and \lam1354.1 was used to constrain the temperature of
the dark voids. A further use of \lam1354.1 from SUMER data was to study
Doppler shift oscillations in hot active region loops
\citep{kliem02,wang03b}, which are triggered by microflares and are
interpreted as standing slow-mode magnetoacoustic waves \citep{wang03a}.

The \ion{Fe}{xxi} line was first reported in a non-solar spectrum by
\citet{maran94} who measured it in a Goddard High Resolution
Spectrometer spectrum of the star AU Microscopii. A survey of
measurements of the line in cool stars was presented by
\citet{ayres03} based on 
 Space Telescope
Imaging Spectrograph (STIS) data.

\ion{Fe}{xxi} is only expected to be found in the solar spectrum
during flares and there are two key spatial locations expected
from the standard model of solar flares. The standard model
\citep[see the review of][]{benz08} specifies an energy release site in the
corona, usually assumed to be the location of magnetic
reconnection. Energy from the release site is transmitted down the
legs of coronal loops in the form of non-thermal particles, a thermal
conduction front, or Alfv\'en waves. The energy is deposited in the
dense chromosphere, leading to plasma heating, and the subsequent
expansion of the plasma results in the ``evaporation'' of hot plasma
into the coronal loops. \ion{Fe}{xxi} is formed at $\approx 10$~MK,
and we expect to find 10~MK plasma concentrated at the loop footpoints due to the
initial chromospheric heating, followed later by the appearance of
bright 10~MK loops that are formed from the evaporated plasma. A key
prediction of the standard flare model is 
that the plasma at the footpoints will flow upwards into the loops as
the evaporation process proceeds and thus blue-shifted emission lines
are expected. Spatially-resolved spectral data from the Coronal
Diagnostic Spectrometer (CDS) on board SOHO, and the EUV Imaging Spectrometer (EIS) on board
the Hinode spacecraft have demonstrated that blue-shifted emission is
seen in emission lines of \ion{Fe}{xix}, \ion{Fe}{xxiii} and
\ion{Fe}{xxiv} and speeds can be as high as 400~\kms\
\citep{2003ApJ...588..596T, 2006A&A...455.1123T, 2006ApJ...638L.117M,
  delzanna06, 
  watanabe11,young13}. The flare loop footpoints are expected to coincide
with the flare ribbons that are seen in H$\alpha$ and  UV continuum
images, although this can not be directly confirmed with CDS and
EIS, which observe extreme ultraviolet radiation.

The Interface Region Imaging Spectrometer (IRIS) was launched in
2013~June and it observes the 1332--1358~\AA\
wavelength region at  a spatial
resolution of 0.33--0.40\as, significantly better than the previous
instruments that have observed the \ion{Fe}{xxi} line. The high
spatial resolution enables the fine-scale 
structure of the \ion{Fe}{xxi} line to be investigated, particularly
with regard flare loops and their footpoints. 
The first study of the
\ion{Fe}{xxi} line using IRIS data was performed by \citet{tian14} who
analyzed the C1.6 class flare that peaked at 17:19~UT on 2014 April
19. The authors reported a red-shifted component that could be
attributed to the downward reconnection outflow expected from the
standard solar flare model.
In this paper we present observations
obtained by IRIS of the X1 class flare that occurred on 2014 March 29,
peaking at 17:48~UT (SOL2014-03-29T17:48). 
\citet{judge14} investigated
photospheric heating and the causes of the sunquake associated with
this flare using data from the Facility Infrared Spectrometer at the Dunn Solar
Telescope. The Balmer continuum increase associated with the flare was
presented by \citet{heinzel14} who used the near-UV continuum
increase measured by IRIS around 2800~\AA. The filament eruption
associated with the flare was studied by Kleint et al. (2014, ApJ, in
press), who combined multiple data-sets, including from IRIS, to measure
the rapid acceleration of the filament, which reached speeds of up to 600~\kms.
Li et al.~(2014, ApJ, in press)
combined data from IRIS and EIS to study evaporation flows at specific
sites in the active region. They demonstrated that the
high-temperature evaporation
flows move with time in response to the flare ribbon separation, and
find redshifts of transition region lines at the locations of the
blue-shifted hot lines (10~MK). The present work differs from this
latter work in studying all of the IRIS ribbons sites, and
investigating the connection to the chromospheric flare ribbons. In
addition we also study the \ion{Fe}{xxi} emission from the post-flare
loop arcade.

The paper is structured as follows. Sect.~\ref{sect.obs} gives an
overview of the March 29 flare observation and describes the
data-sets. Sect.~\ref{sect.ribbons} defines flares ribbons and kernels
in the context of the IRIS data, and Sect.~\ref{sect.fe21r} describes
how the \ion{Fe}{xxi} emission in the vicinity of the ribbons
behaves. Sect.~\ref{sect.fe21l} discusses properties of the post-flare
loop arcade, and Sect.~\ref{sect.summary} summarizes the results.

\section{Observations}\label{sect.obs}

\begin{figure}[h]
\epsscale{0.7}
\plotone{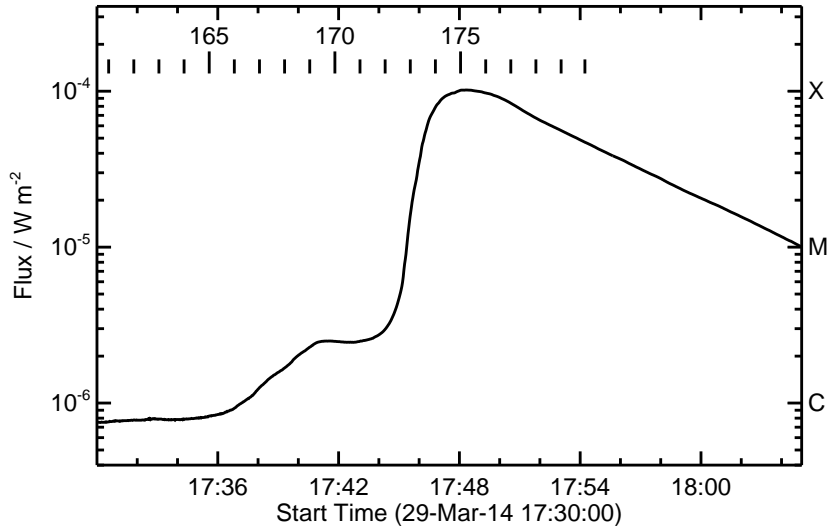}
\caption{The GOES 1--8~\AA\ light curve for the period 17:30--18:05~UT
on 2014 March 29. Short vertical lines indicate the start and end
times of the IRIS rasters, and the IRIS raster numbers are indicated.}
\label{fig.goes}
\end{figure}

The flare occurred in active region AR 12017 which was part of a
larger complex that included AR 12018. The 1--8~\AA\ GOES light curve
is plotted in Figure~\ref{fig.goes} and shows that the flare began at 17:35~UT
with an initial rise from a GOES class level B9 to C3 by 17:41~UT. A
rapid rise to the X1 level began at 17:44~UT, with the peak reached at
17:48~UT. 

IRIS  is described in detail by \citet{iris} and is
briefly summarized here. The instrument returns narrow-slit spectral
data over three wavelength bands of 1332--1358, 1389--1406 and
2782--2834~\AA, and a slit-jaw imager can produce images centered at
1330, 1400, 2796 and 2832~\AA, giving context images for the
spectra. The spectrometer slit is 0.33\as\ wide, and the detector
pixel sizes correspond to 0.167\as. The strongest emission lines
observed by the instrument are \ion{Mg}{ii} \lam\lam2796.4, 2803.5,
\ion{C}{ii} \lam\lam1334.5, 1335.7, and \ion{Si}{iv} \lam\lam1393.8,
1402.8, which are formed in the chromosphere and transition
region. \ion{Fe}{xxi} \lam1354.1 is the only coronal line in the
present data-set that can
yield sufficient signal-to-noise for detailed studies of line
profiles. 

For the present flare observation IRIS performed 180 raster scans over
the period 14:09 to 
17:54~UT. Each scan consisted of eight slit positions separated by
2\as, giving a total raster size of 14.33\as. 
The field-of-view in the
solar-Y direction was 175\as\ and 8~second exposures were used. The
duration of each raster was 75~seconds. A
slit-jaw image was obtained with each raster exposure: 
images at 2796~\AA\ were obtained for exposures 1, 3, 5 and 7;
images at 1400~\AA\ were obtained for exposures 0, 4 and 6; and an
image at 2832~\AA\ was obtained for exposure 2.
Saturation on the detector was a problem for the strong IRIS lines
during the flare (particularly \ion{Si}{iv} \lam1402), but \ion{Fe}{xxi} \lam1354.1
was not saturated at any location. We focus on rasters 171--179, which
cover the rise and initial-decay phases of the flare
(Figure~\ref{fig.goes}). We note that there was weak \ion{Fe}{xxi}
from the active region prior to this time period that is related to
the weaker C3 event, but our focus in the present work is on the flare
ribbons and post-flare loops of the X1 event.
As a shorthand in the present work we will refer to specific IRIS
exposures as, e.g., R174E0, which refers to raster no.~174 and
exposure no.~0. The raster numbers go from 0 to 179 and the exposure
numbers from 0 to 7 (note that IRIS rasters from east to west). When
giving times for 
these exposures, we give the mid-time of the exposure.

Level-2 IRIS data were downloaded from the IRIS website and they are corrected
for most instrumental effects, including geometric distortions,
flat-fields and spatial alignments. The version of the calibration
processing routine (IRIS\_PREP) applied to the data was
1.27. Intensities are given in data-number 
(DN) units and 1-$\sigma$ uncertainties are derived as follows. For
the FUV channel, one DN corresponds to four detected photons
\citep{iris}, so $d$~DN yield $p=4d$ photons. Assuming Poisson noise,
the uncertainty on $p$ is $\sqrt{p}$, however the dark current
uncertainty, $\sigma_{\rm dc}$, also needs to be included, giving
\begin{equation}
\sigma_p= \sqrt{4d + (4\sigma_{\rm dc})^2}
\end{equation}
\citet{iris} give a value of $\sigma_{\rm dc}=3.1$~DN for the FUV
channel. Defining the fractional error to be $f=\sigma_p/p$, the uncertainty on
$d$ is then $\sigma_d=fd$. This method for computing the intensity
uncertainties is implemented in the \emph{Solarsoft} IDL routine IRIS\_GETWINDATA.

We found  that the best way to study  the \lam1354.1 emission line for
this data-set was
through exposure images (wavelength vs.\ solar-Y) as they allow  the
difference in morphology between the broad 
\lam1354.1 line and the narrow blending cool lines to be more easily discerned.
The raster is too
sparse for X-Y images to accurately reflect the intensity
distribution across the raster and for this it is better to consider
131~\AA\ images from the Atmospheric Imaging Assembly on board the
Solar Dynamics Observatory. 

The AIA instrument is described in \citet{aia}, and the images
considered in the present work are  from 
the 131~\AA\ filter  (hereafter referred to as ``A131''), which is
dominated by \ion{Fe}{xxi} \lam128.75 during flares 
\citep{odwyer10} and thus
is ideal for comparisons with the IRIS slit images. Additional
contributing lines are \ion{Fe}{xx} \lam132.84 and \ion{Fe}{xxiii}
\lam132.91. The image spatial scale is 0.6\as\ per pixel, and the
cadence is 12~seconds.
An automatic observing procedure is initiated during
flares when count rates become large. Every alternate frame remains at
the nominal exposure time, with the exposure times for the remaining
frames adjusted through an automatic exposure control
mechanism. Despite this, the rapid intensity increases during the
flare still led to many frames that were badly affected by saturation.
In the present case from  17:46~UT to the end of the IRIS sequence at
17:54:19~UT there are only 12  A131 
exposures for which saturation is sufficiently low to allow comparison
with the IRIS data, however these images still yield valuable comparisons with
the IRIS data as demonstrated in the following sections.

The wavelength of \ion{Fe}{xxi} \lam1354.1 was given as $1354.08\pm 0.05$~\AA\ by
\citet{sandlin77} based on \emph{Skylab} S082B
measurements, and \citet{feldman00} gave a value of $1354.064\pm
0.020$~\AA\ from SUMER spectra. In the present article we derive a new
value for the reference wavelength of $1354.106\pm 0.023$~\AA\
(Appendix~\ref{sect.ref}). 
The temperature of maximum emission of \lam1354.1 is $\log\,T/{\rm K}=7.06$, as
obtained from atomic data in version 7.1 of the CHIANTI database
\citep{dere97,chianti71}. This temperature translates to a full-width at
half-maximum (FWHM) emission line width of 0.438~\AA, or 96.9~\kms\ in
velocity space. The instrumental
width is 25.85~m\AA\ \citep{iris} and so is essentially negligible when
studying \lam1354.1 line width measurements.

\section{Flare ribbons}\label{sect.ribbons}

As stated in the Introduction, energy from the coronal energy
release site of a flare travels to the chromosphere down coronal loop
legs, leading to heating that is revealed through bright flare
ribbons seen in H$\alpha$ and UV continuum images. We expect to see
\ion{Fe}{xxi} emission at these ribbons due to the rapid heating that
occurs there. In this section we discuss properties of flare ribbons
and their appearance in IRIS data.

Flare ribbons are a characteristic feature of H$\alpha$ flare
observations and they appear as narrow, curved, bright lines across the
solar surface. Pairs of ribbons are common, but they can have complex
shapes and discontinuities. The ribbons resolve into lines of compact
kernels, and the sizes of the kernels have been measured to
0.6--1.9\as\ in one case, with smaller sizes corresponding to deeper
atmospheric depths \citep{xu12}. Flare kernels can be seen in white
light flares, but are much more easily seen in the UV continuum, such
as the 1700~\AA\ channel of the Transition and Coronal Explorer
\cite[TRACE,][]{trace} and AIA. \citet{young13} reported an 
AIA measurement of a flare kernel and found that it emitted in all of the
the EUV channels of AIA, with a size at the resolution of the
instrument ($\approx 1.2$\as). 

The IRIS instrument observes at FUV and near-ultraviolet (NUV)
wavelengths, and 
the ribbons can be measured in both the
continuum and the emission lines. The FUV continuum enhancement during
flares is believed to be
driven by backwarming from \ion{C}{ii} \lam\lam1334, 1335 and
\ion{Si}{iv} \lam\lam1393, 1402 \citep{doyle92}. We thus expect the
FUV continuum ribbons to be correlated with the brightness of these
emission lines (which are measured by IRIS). Although beyond the scope
of the present work, we note that inspection of the March 29 flare
suggests that this is indeed the case. 

\begin{figure}[h]
\epsscale{1.0}
\plotone{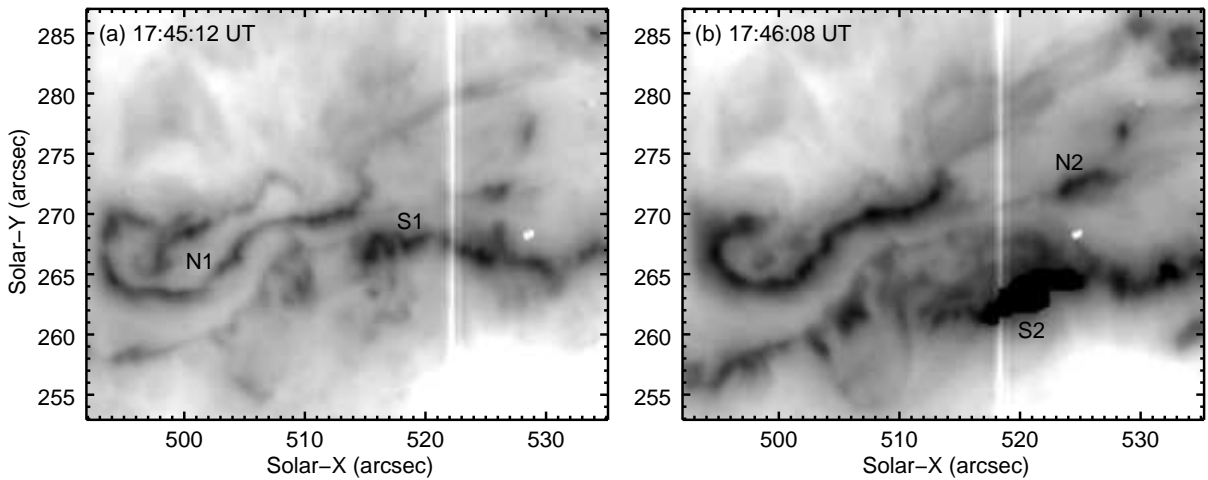}
\caption{IRIS 2796~\AA\ slit-jaw images at two different times during
  the development of the flare ribbons. An inverse logarithmic
  intensity scaling is used. }
\label{fig.2796}
\end{figure}

The development of the March~29 ribbons is best studied with the IRIS
slit-jaw data, in particular the 2796~\AA\ images that were obtained
at an 18~second cadence. These images show significantly less
saturation than the 1400~\AA\ images, allowing the ribbon evolution to
be seen more clearly. The 2796~\AA\ images are dominated by emission
from \ion{Mg}{ii} and we note that the ribbon structures are less
well-defined than in the 1400~\AA\ images, although this may just
reflect the lower spatial resolution of the longer wavelength channel \citep{iris}.
 Movie~1 shows the development of the ribbons
over the period 17:42:05 to 17:49:15~UT, and Figure~\ref{fig.2796}
shows images from this movie at 17:45:12 and 17:46:08~UT. The initial
impression from the movie is that 
there are two ribbons: a north ribbon (N1) between X=505 and 515, and a
south ribbon (S1) that extends across the full X-range of the images in
Figure~\ref{fig.2796}. The north ribbon at 17:45:12~UT is seen to double back on
itself, with the two arms being separated by only 2--4\as. The arms
move towards each other slightly before the northerly arm fades, leaving the
southerly arm with a distinctive hook on the east side at around
X=495 (Figure~\ref{fig.2796}b). In addition to this ribbon, there is a
small ribbon that is labeled N2 in Figure~\ref{fig.2796}b and is
bright in exposures E6 and E7 of the rasters. There is a faint, eastward
extension of this ribbon that possibly connects to N1.

The south ribbon becomes much brighter than the north ribbon,
with some saturation in the 2796~\AA\ images. It also moves southwards
by about 10\as\ from 17:45:12~UT to the end of the movie at
17:49:15~UT. Between X=510 and 525, the southward movement of the
ribbon leaves behind a patch of enhanced emission and the IRIS spectra
demonstrate that the FUV continuum is enhanced in this region. The
spectra also reveal that the initial position of the south ribbon
(marked with S1 in Figure~\ref{fig.2796}a) maintains enhanced
continuum and emission lines even as the main ribbon (S2) advances
southwards. From Movie~1 it is not clear if S2 actually represents the
new position of S1 or whether it is a distinct ribbon. A crucial
1400~\AA\ 
slit-jaw image at 17:45:26 UT that may resolve this question is too
saturated to allow the south ribbon to be identified. We  distinguish S1 from S2 based on the fact that it retains a
distinctive signature in terms of enhanced continuum and emission
lines.

In the following section we are interested in the location of \ion{Fe}{xxi} emission relative
to the ribbons, and we define the ribbons in the spectral images as
being the locations where the
FUV continuum is enhanced. Comparison of 1400~\AA\ and 2796~\AA\
slit-jaw images, and FUV and NUV spectral images suggests that the FUV
continuum ribbons directly correspond to the 2796~\AA\ image ribbons
shown in Movie~1.

\section{Fe XXI at the ribbons}\label{sect.fe21r}

In this section we interpret the \ion{Fe}{xxi} emission close to the
flare ribbon sites in terms of the standard flare model, which
invokes heating of the chromosphere at the
footpoints of an arcade of coronal loops to explain the bright
chromospheric flare ribbons. The plasma is heated to temperatures of
$\sim 10$~MK, and it rises up the loops, filling them with hot, dense
plasma to create the bright post-flare loop arcade. The loop
footpoints will emit in \ion{Fe}{xxi} \lam1354.1 and we expect to see
blue-shifts from the evaporating plasma. 
Key questions that will be addressed are: (i) does \ion{Fe}{xxi}
always appear at the chromospheric ribbon sites? (ii) is \ion{Fe}{xxi}
alway blue-shifted at the ribbons?  (iii) is there a delay between the
chromospheric ribbon appearing and \ion{Fe}{xxi} being produced? and
(iv) where the position of the ribbon moves in time (implying new sets
of nested loops being progressively heated) do we continue to see
blue-shifted \ion{Fe}{xxi} emission at the older sites?

\begin{figure}[h]
\epsscale{0.8}
\plotone{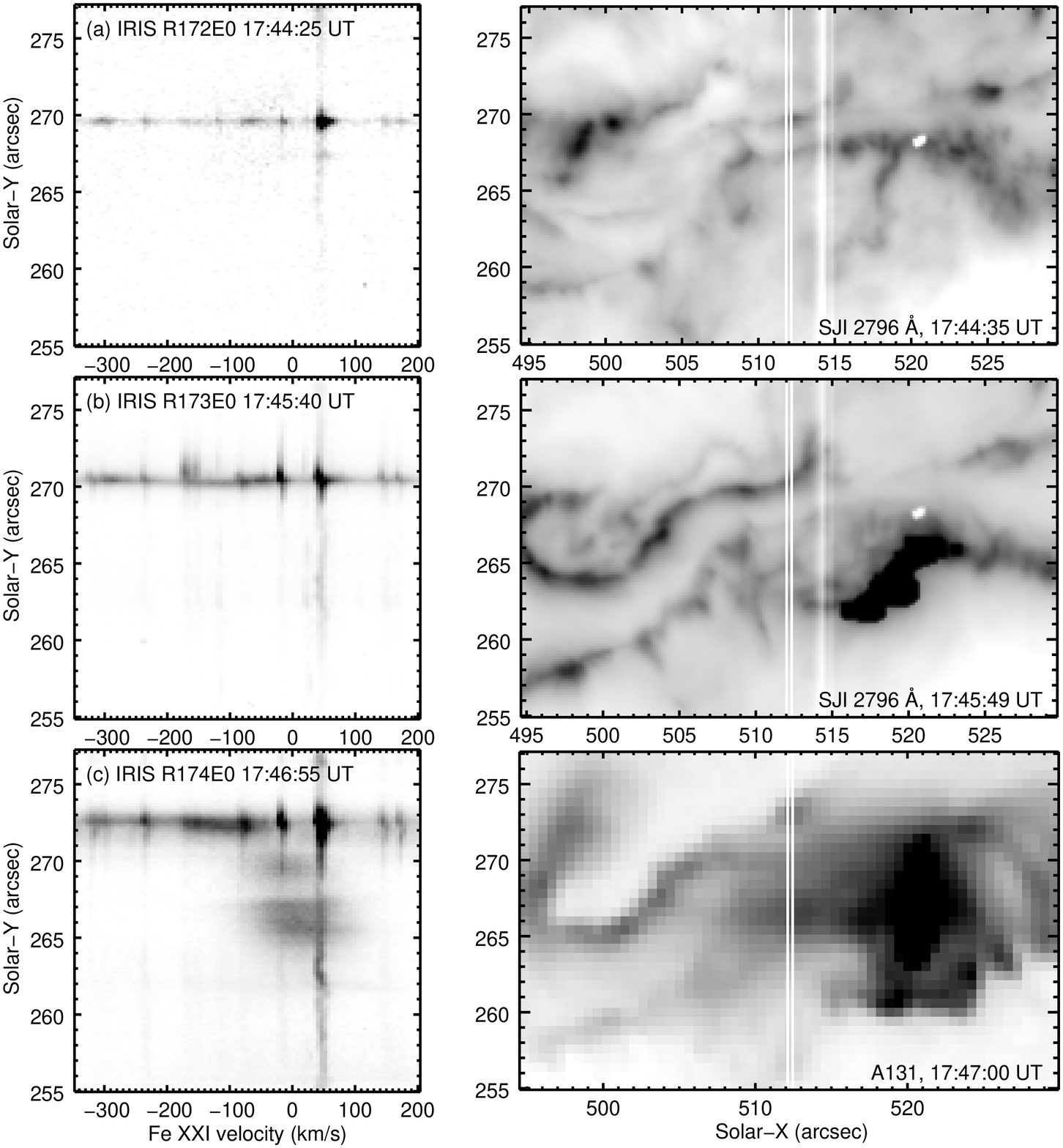}
\caption{Sequence of three images showing the development of the
  \ion{Fe}{xxi} \lam1354.1 line at a location on the N1 flare
  ribbon. The left column shows IRIS detector images with a
  reverse--linear intensity scaling applied. A saturation has been
  applied to better reveal the weak \ion{Fe}{xxi} line. The right
  columns of panels a and b show IRIS 2796~\AA\ slit-jaw images, with
  the location of the IRIS slit corresponding to the left column
  images indicated by white parallel lines. An inverse--logarithmic
  intensity scaling has been applied. The right column of panel c
 shows an A131 image for which an
  inverse--square-root intensity scaling has been applied. The
  parallel white lines indicate the location of the IRIS slit.}
\label{fig.e0}
\end{figure}

\begin{figure}[h]
\epsscale{0.9}
\plotone{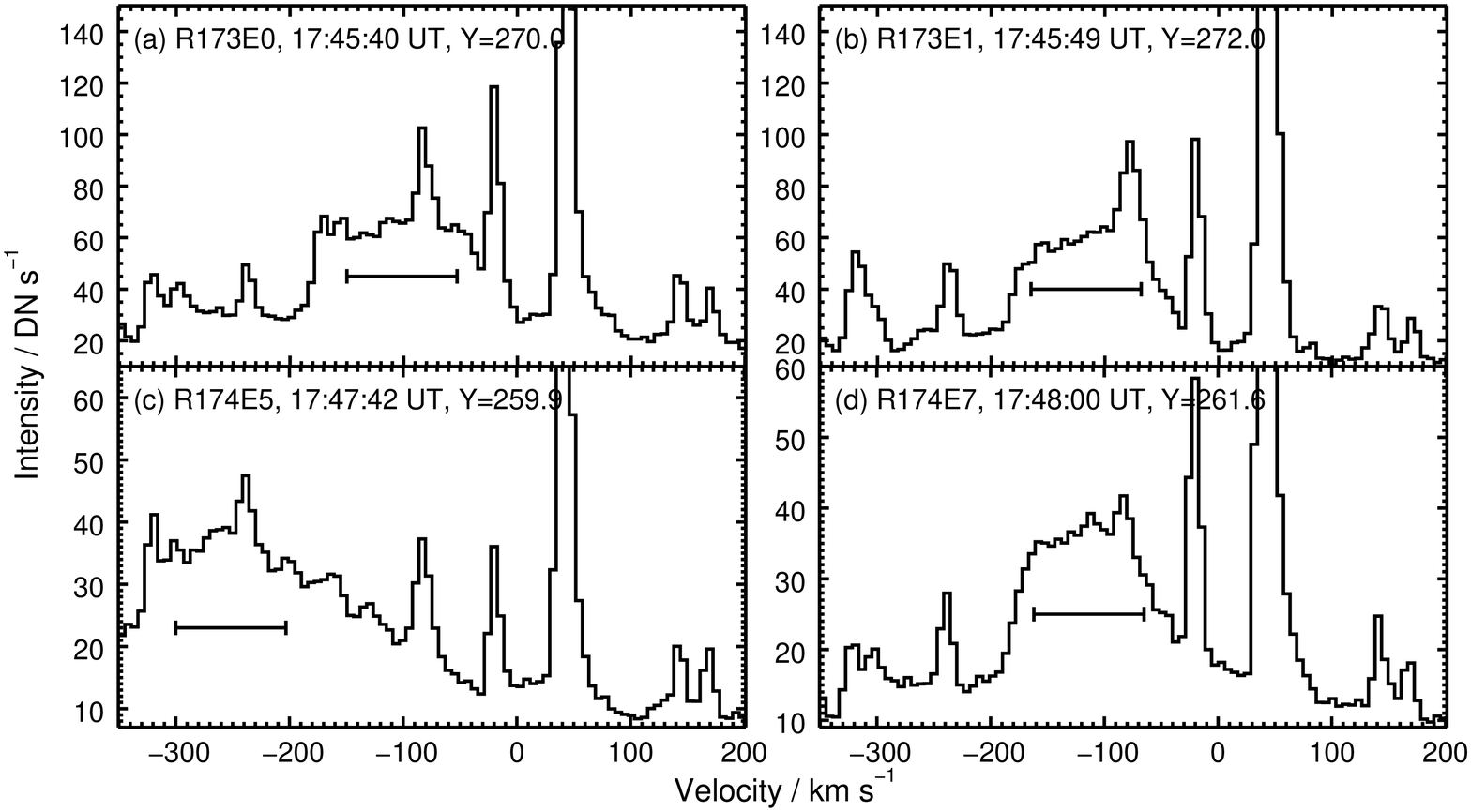}
\caption{Four examples of \ion{Fe}{xxi} \lam1354.1 line profiles from
  locations at or close to the flare ribbons. The profiles have been
  obtained for individual solar-Y pixels, and the Y-values are
  indicated on the plots. The horizontal lines show the thermal width
  expected for $\log\,T=7.06$, the temperature of formation of \ion{Fe}{xxi}.
}
\label{fig.profiles}
\end{figure}

Our first task is classify the spectroscopic observations of the four ribbons (N1,
N2, S1 and S2), and Table~\ref{tbl.ribbon} presents the results. By
inspecting the sequence of rasters we identify the time at which the
ribbon appears in each of the eight raster exposures. For example, the
N1 ribbon is only observed in exposures E0 and E1, and it first
appears in E0 during raster R172 (17:44:25~UT). This is illustrated in
Figure~\ref{fig.e0} where panel a shows the FUV continuum and
chromospheric lines brighten over a small spatial region of width only
0.3\as\ in the solar-Y direction. There is no \ion{Fe}{xxi} emission
in this exposure. The next raster, R173 (Figure~\ref{fig.e0}b), shows
that the continuum and chromospheric lines have brightened and now
\ion{Fe}{xxi} is present, slightly offset in Y from the continuum, although it is
weak. Figure~\ref{fig.profiles}a shows a 1D spectrum obtained at
Y=270.0\as, offset 0.33\as\ from the ribbon to better show the
\ion{Fe}{xxi} line. It is clearly seen to be broad and 
blue-shifted by around
100~\kms. Figure~\ref{fig.e0}b shows that the \ion{Fe}{xxi} emission
is very compact in the Y-direction (around 0.3\as) and slightly offset
from the location of the ribbon emission. The ribbon moved slightly
northwards between rasters R172 and R173, and so the offset may
indicate that \ion{Fe}{xxi} is emitted from the earlier location of
the ribbon. Figure~\ref{fig.e0}c shows the next raster with a clearly
visible \ion{Fe}{xxi} line at the ribbon site, and two additional
patches of \ion{Fe}{xxi} emission at 266\as\ and 270\as\ that can be
identified as loop emission due to their broader extent in the
Y-direction and their small Doppler shifts. All three \ion{Fe}{xxi}
locations can be identified in the co-temporal A131 image,
demonstrating that this image mostly displays \ion{Fe}{xxi} at this
stage of the flare.

\begin{deluxetable}{cccccccc}
\tablecaption{Properties of the IRIS flare ribbons.\label{tbl.ribbon}}
\tablehead{
  &&& Ribbon & \ion{Fe}{xxi} & Velocity \\
  Ribbon &Exposure & Solar-Y & appears\tablenotemark{a} & appears\tablenotemark{a}  & (\kms)
  &Position & Morphology\tablenotemark{b}
}
\startdata
N1 & 0 & 269.6 & 44:25 [172] & 45:40 [173] & $-100$ & Offset 
       & Compact \\
   & 1 & 272.1 & 45:49 [173] & 45:49 [173] & $-100$ & Ribbon 
       & Compact \\
\noalign{\smallskip}
N2 & 3 & 270.4 & 46:08 [173] & 46:08 [173] & $0$ & Ribbon 
       & Extended (N) \\
   & 4 & 271.1 & 46:17 [173] & 46:17 [173] & $+30$ & Ribbon 
       & Extended (N,S) \\
   & 5 & 271.2 & 46:27 [173] & 46:27 [173] & $0$ & Ribbon 
       & Compact \\
   & 6 & 271.4 & 44:06 [171] & 46:36 [173] & $-120$ & Ribbon 
       & Diagonal-in (S) \\
   & 7 & 271.3 & 45:31 [172] & 46:45 [173] & $-10$ & Ribbon 
       & Extended (N,S) \tablenotemark{c}\\
\noalign{\smallskip}
S1 & 1 & 268.3 & 45:49 [173] & 47:04 [174] & $0$ & Ribbon 
       & Extended (N,S)\tablenotemark{c}\\
   & 2 & 268.3 & 44:44 [172] & 45:59 [173] & $-20$ & Ribbon 
       & Extended (S) \\
   & 3 & 267.9 & 44:53 [172] & 46:08 [173] & $-30$ & Offset 
       & Extended (S) \\
   & 4 & 268.2 & 45:03 [172] & 46:17 [173] & $-10$ & Ribbon 
       &Extended (S)\tablenotemark{c} \\
   & 5 & 267.2 & 45:12 [172] & 46:27 [173] & $0$ & Ribbon 
       & Extended (S)\tablenotemark{c}\\
\noalign{\smallskip}
S2 & 1 & 262.2 & 45:49 [173] & 47:04 [174]  & $-100$ & Ribbon 
       & Diagonal-in (N) \\
   & 2 & 261.8 & 45:59 [173] & 47:14 [174] & $-120$ & Offset 
       & Diagonal-in (N) \\
   & 3 & 262.5 & 46:08 [173] & 47:23 [174]  & $-220$\tablenotemark{d} 
       & Ribbon & Diagonal-in (N) \\
   & 4 & 262.4 & 46:17 [173] & 46:17 [173] & $-120$ & Offset 
       & Compact \\
   & 5 & 261.8 & 46:27 [173] & 47:42 [174] & $-160$ & Offset 
       & Diagonal-out (N) \\
   & 6 & 266.1 & 45:21 [172] & 45:21 [172] & $-20$ & Offset 
       & Diagonal-out (N) \\
   & 7 & 265.3 & 45:31 [172] & 48:00 [174] & $-110$ & Offset 
       & Compact \\
\enddata
\tablenotetext{a}{All times are 17:MM:SS, where MM:SS are
  indicated. Square brackets denote the IRIS raster number.}
\tablenotetext{b}{N or S in brackets indicates whether the emission
  extends northwards or southwards from the ribbon site.}
\tablenotetext{c}{Contaminated with flare loop emission.}
\tablenotetext{d}{Very broad and low intensity, so velocity uncertain.}
\end{deluxetable}

We repeat this analysis for all of the other ribbon locations, giving
the results in Table~\ref{tbl.ribbon}. In several locations
interpretation is not easy, for example due to contamination by
overlying \ion{Fe}{xxi} loop emission which is identified by emission
close to the rest wavelength of the line, and by a greater spatial
extent in the solar-Y direction. There are also cases where there are
two ribbons very close to each other that we interpret as a kink in
the ribbon in which the slit crosses the ribbon twice. Examples
include R173E5 and R172E7, and evidence for the kinked ribbons can be
seen in Movie~1. We choose the brighter ribbon location in these
cases.

Figures~\ref{fig.e0} and \ref{fig.profiles}a demonstrate that a number
of chromospheric lines become strong at the ribbon location,
compromising studies of the \ion{Fe}{xxi} line. The examples from the
March~29 flare show that the blending lines are always present at the
ribbons, although they are stronger for the S2 ribbon compared to the
other ribbons. A further complication is that some of the lines,
particularly, \ion{Si}{ii} \lam1352.64 and \lam1353.72 (at velocities
$-325$ and $-85$~\kms\ relative to \lam1354.1), can exhibit
significant broadening and Doppler shifts in some locations. 
Further examples of ribbon
spectra are shown in Figure~\ref{fig.profiles}, and a discussion of
the blending lines is given in Appendix~\ref{sect.blend}.

The results for the 19 ribbon sites firstly confirm that all of the
ribbon locations can be identified with co-spatial or slightly offset
\ion{Fe}{xxi} emission, although the emission line is not always
blue-shifted. In general there is a lag
of one raster (75~seconds) between the first appearance of the
chromospheric ribbon and the first appearance of \ion{Fe}{xxi} at, or
close-to the ribbon site.

Ribbons N1 and S2 are the brightest ribbons and they fit the classical
picture of two approximately parallel ribbons located either side of the
polarity inversion line that separate as the flare progresses. Of the
nine sites along these ribbons, eight of them show blue-shifted
\ion{Fe}{xxi} emission consistent with the chromospheric evaporation
scenario. Ribbons N2 and S1 are different in that they are dimmer and
show little or no motion with time (Movie~1). Only one location
(R173E6) shows a large \ion{Fe}{xxi} blueshift, even though the line
can be clearly identified with the chromospheric site. The smaller
blueshifts may indicate that gentle evaporation is occurring at these
sites, although there is one location where a redshift is found
(ribbon N2, exposure 4).

\begin{figure}[h]
\epsscale{1.0}
\plotone{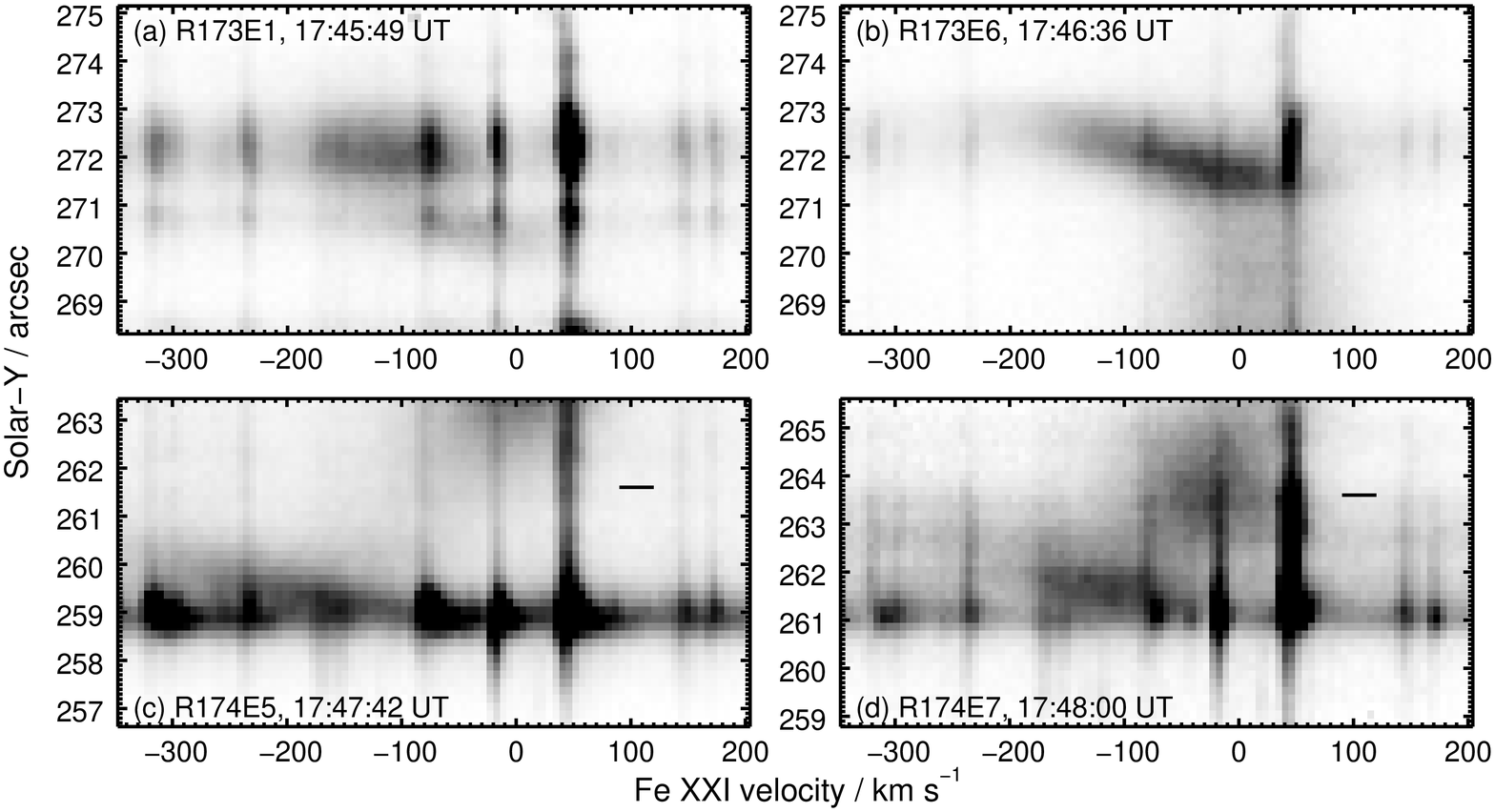}
\caption{Four IRIS detector images showing different examples of
  \ion{Fe}{xxi} emission at the flare ribbon sites. An inverse-linear
  intensity scaling has been applied, and the images have been
  saturated to better reveal the \ion{Fe}{xxi} line. Short horizontal
  lines on panels c and d show the position of the flare ribbon in the
previous rasters. }
\label{fig.closeups}
\end{figure}

In Table~\ref{tbl.ribbon} we list the position and morphology of the \ion{Fe}{xxi} emission as
seen on the detector images.
The position can either be
coincident with the chromospheric location, or offset from it. The
offsets are small, $\le 1$\as, but significant at the IRIS spatial
resolution. Note that the majority of offset \ion{Fe}{xxi} emission is
associated with the fast-moving S2 ribbon. The morphology of the
\ion{Fe}{xxi} emission is expressed according 
to the spatial extent in the Y-direction. Compact and extended
indicate whether the emission extends less than or more than 1\as,
while ``diagonal-in'' and ``diagonal-out'' refer to extended emission
that changes position in the wavelength direction with
position. Examples are shown in Figure~\ref{fig.closeups}.

The interpretation of the extended, diagonal-in and diagonal-out
emission types is not straightforward. For example, if the IRIS slit
is aligned along the leg of one of the post-flare loops, then we may
expect to see a large blue-shift of \ion{Fe}{xxi} at the location of
the flare ribbon, corresponding to evaporating chromospheric
plasma. Further up the loop leg we may expect the magnitude of the
blueshift to decrease as the velocity of the upflowing plasma
decreases. This may explain the morphology of \ion{Fe}{xxi}
from the N2 ribbon for which the blueshift
decreases from over 100~\kms\  to 0~\kms\ over less than 2\as\
(Figure~\ref{fig.closeups}b) -- one
of the diagonal-in types.

However, we know that the N1 and S2 ribbons change position with time,
and so examples such as Figure~\ref{fig.closeups}c and d --
diagonal-out and offset-extended types, respectively -- may take their
appearance because \ion{Fe}{xxi} is emitted from an earlier location
of the ribbon. In Figure~\ref{fig.closeups}c and d we have indicated
the locations of the chromospheric ribbon in the previous rasters
(obtained 75~seconds earlier) with a short horizontal line,
demonstrating that the ribbon moved by around 2.5\as\ in each case.
Figure~\ref{fig.closeups}c shows larger
blueshifts about 0.5--1.0\as\ away from the ribbon site, so this may
indicate that the blueshift at a fixed spatial location increases
sometime after the initial burst of chromospheric heating that causes
the chromospheric ribbon. Figure~\ref{fig.closeups}c does not show
such a striking velocity pattern, but the brightest \ion{Fe}{xxi}
emission occurs about 0.5\as\ from the emission, and there is
quite bright \ion{Fe}{xxi} emission at the earlier location of the
ribbon, although with a velocity only around $-20$~\kms. Close
inspection of both the R174E5 and R174E7 locations does show that
\ion{Fe}{xxi} is present at all Y-locations between the previous and
new ribbon
locations, even though the brightness is variable. 

What is clear from the fast-moving  S2 ribbon is that the ribbon does
\emph{not} leave behind a trail of uniformly bright and blue-shifted \ion{Fe}{xxi} at the
former locations of the ribbon. Instead there is patchy
emission that is usually brightest close to or at the current ribbon
site, and that can have a range of velocities. This may reflect the
non-uniform  process that heats the chromosphere.

\begin{figure}[h]
\epsscale{0.5}
\plotone{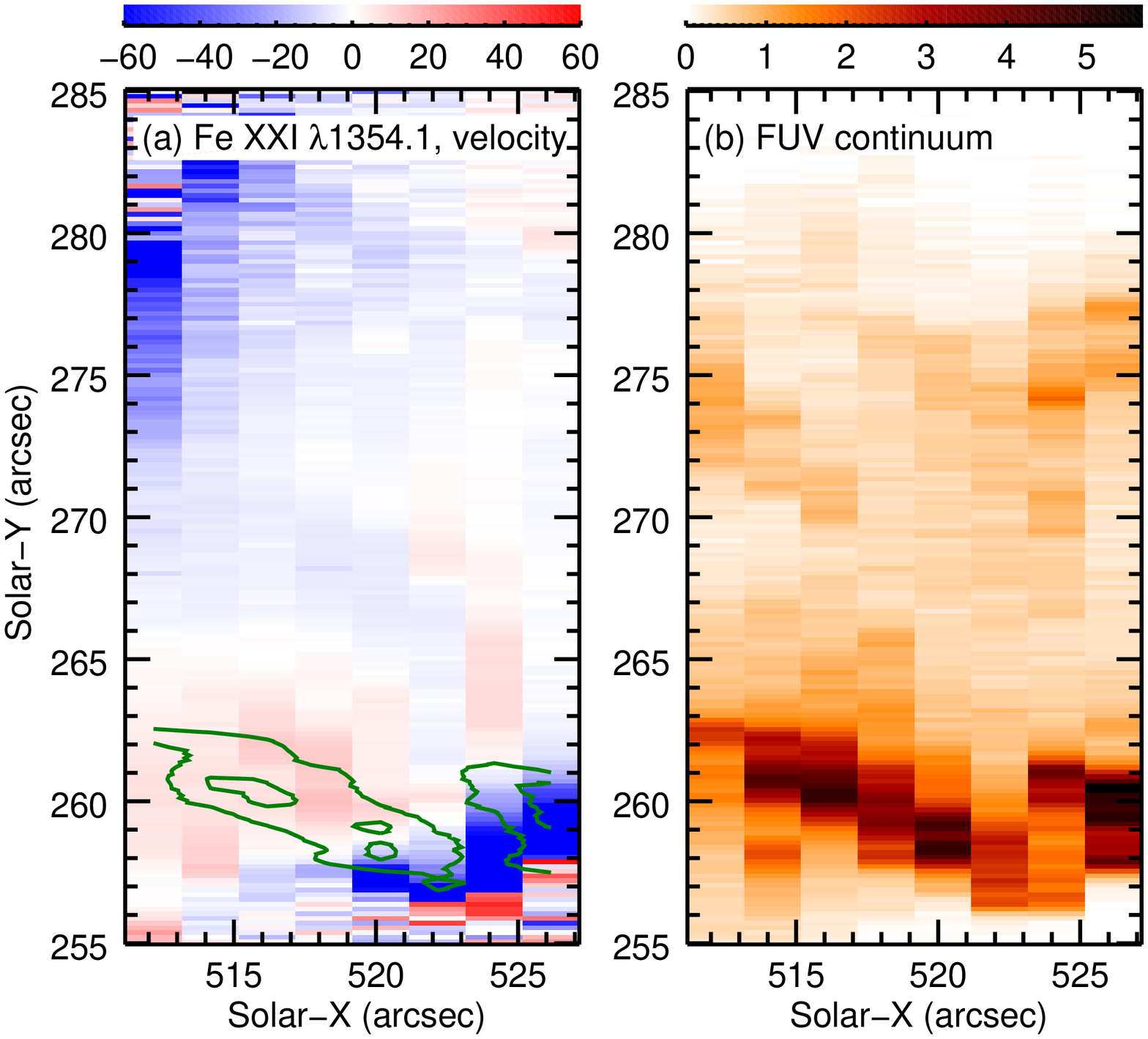}
\caption{The left panel shows the LOS velocity of \ion{Fe}{xxi}
  \lam1354.1 from IRIS raster R179 (17:53:05--17:54:20
  UT), as derived from a Gaussian fit. The right panel shows the FUV
  continuum level neighboring the \ion{Fe}{xxi} line, and the green
  contours on panel a show the continuum level at 2.5 and 4.5~DN.}
\label{fig.late}
\end{figure}

The IRIS observations terminate at 17:54:19~UT, only 6~minutes after
the flare peak, and blue-shifted \ion{Fe}{xxi} is still present at
this time as demonstrated in Figure~\ref{fig.late}. This figure shows
the LOS velocity of \lam1354.1 and the FUV continuum level as
determined from fits to the IRIS spectra in the 1352.3--1355.9~\AA\
region -- see Sect.~\ref{sect.fe21l}. The S2 ribbon is still visible
through an enhanced continuum level, and \lam1354.1 is blue-shifted by
20--60~\kms\  in exposures E4--7 at the location of the ribbon. There
are also blueshifts related to the former position of the N1 ribbon seen in exposures E0--1,
although there is no enhanced continuum at these locations.

To summarize the results in relation to the four questions
posed earlier in this section:
\begin{itemize}
\item \ion{Fe}{xxi} is found at all of the chromospheric ribbon sites.
\item \ion{Fe}{xxi} is not always blue-shifted at the ribbon
  sites. However for the two brightest ribbons (N1, S2) the line is
  blue-shifted at 8 of the 9 sites. The N2 and S1 ribbons seem to be
  anomalous and generally do not show blueshifts.
\item There is generally a lag of 75~seconds (the raster cadence
  time) before the line appears.
\item The S2 ribbon moves quite rapidly over the solar surface, and
  \ion{Fe}{xxi} is found at the previous locations of the
  ribbon. However, the emission is patchy and the magnitudes of the
  blueshifts vary.
\end{itemize}

In addition to these points, we note that the velocities of \ion{Fe}{xxi} \lam1354.1 at the ribbons are generally
quite low, with values typically around $-100$~\kms\ (column 6 of
Table~\ref{tbl.ribbon} and Figures~\ref{fig.profiles}a,b and d),
significantly smaller than the velocities of $\approx -400$~\kms\ found
by \citet{watanabe11} and
\citet{young13} from the \ion{Fe}{xxiii} and \ion{Fe}{xxiv} lines
observed by EIS for two flares. This may be due to lower velocities at
the cooler temperature of \ion{Fe}{xxi} and a less favorable
line-of-sight, although we also note that the IRIS data window used
for the observation did not extend beyond $-350$~\kms\ from the line
center, thus very large velocities $\le -400$~\kms\ could not be
detected. The largest blueshift seen in the IRIS data was from the S1 
ribbon observed in R174E5 (Figure~\ref{fig.closeups}c) for which the
centroid reached $\approx -250$~\kms\ (Figure~\ref{fig.profiles}c). The
profile was also very broad and apparently asymmetric. An asymmetric
profile can also be identified from exposure R174E7
(Figure~\ref{fig.profiles}d).  The blending lines complicate
interpretation of the \lam1354.1 line widths, but the profiles shown
in 
Figure~\ref{fig.profiles} have FWHMs at least as wide as the thermal
width (96.9~\kms), and significantly larger in the case of Figure~\ref{fig.profiles}c.

Another contrasting feature with the profiles presented by \citet{watanabe11} and
\citet{young13} is the lack of a two-component structure to the
\ion{Fe}{xxi} line, in particular the absence of a component near the
rest wavelength. 
This may
suggest that the EIS observations did not resolve the upflowing flare
kernel sites
from the stationary flare loop plasma, although \citet{young13} were
able to demonstrate that the flare kernel they studied was isolated in
co-temporal AIA images. Careful alignment of IRIS and EIS data is
needed to determine whether the stationary component is  missing
from the \ion{Fe}{xxi} line or whether it is simply due to  a blending
of loop and footpoint emission in the EIS data.

To summarize, \ion{Fe}{xxi} \lam1354.1 is not easy to measure at the
flare ribbons on account of the intrinsically weak emission and the
presence of a number of cool emission lines, some of which are
broadened and/or demonstrate enhanced short wavelength wings at the
ribbon site. However, the IRIS data show that \ion{Fe}{xxi} is seen at
or very close to 
all of the ribbon locations, typically about 75~seconds after the ribbons
appear. Blueshifts typically correspond to velocities of $-100$ to
$-200$~\kms, and the spatial extent (in solar-Y) of the \ion{Fe}{xxi} ribbon
emission is about 1\as. A single emission line component is always
seen at the ribbons and the widths are at least as large as the
thermal width of the line. Blue-shifted
\ion{Fe}{xxi} emission at the ribbon locations is clearly seen until
the end of the IRIS sequence at 17:54~UT, six minutes after the flare
peak.

\section{Fe XXI in the flare loops}\label{sect.fe21l}

\begin{figure}[h]
\epsscale{1.0}
\plotone{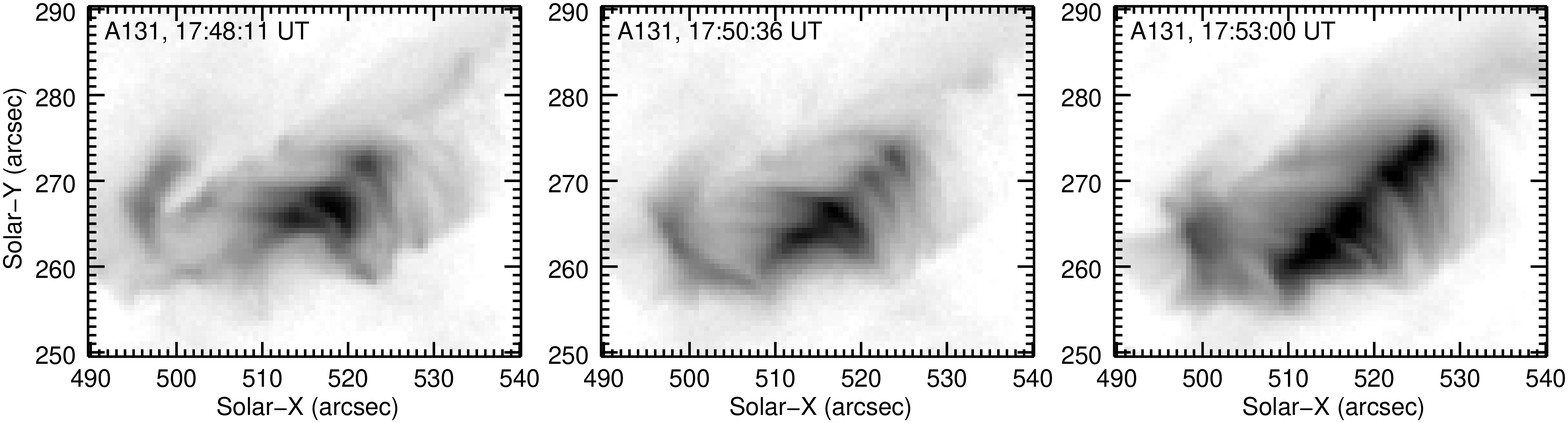}
\caption{Three A131 images showing the post-flare loop arcade at
  different times. A cubed-root scaling has been applied and the color
  table reversed. The brightest areas (black) are saturated in each of
the frames. }
\label{fig.a131-loops}
\end{figure}

\begin{figure}[h]
\epsscale{1.0}
\plotone{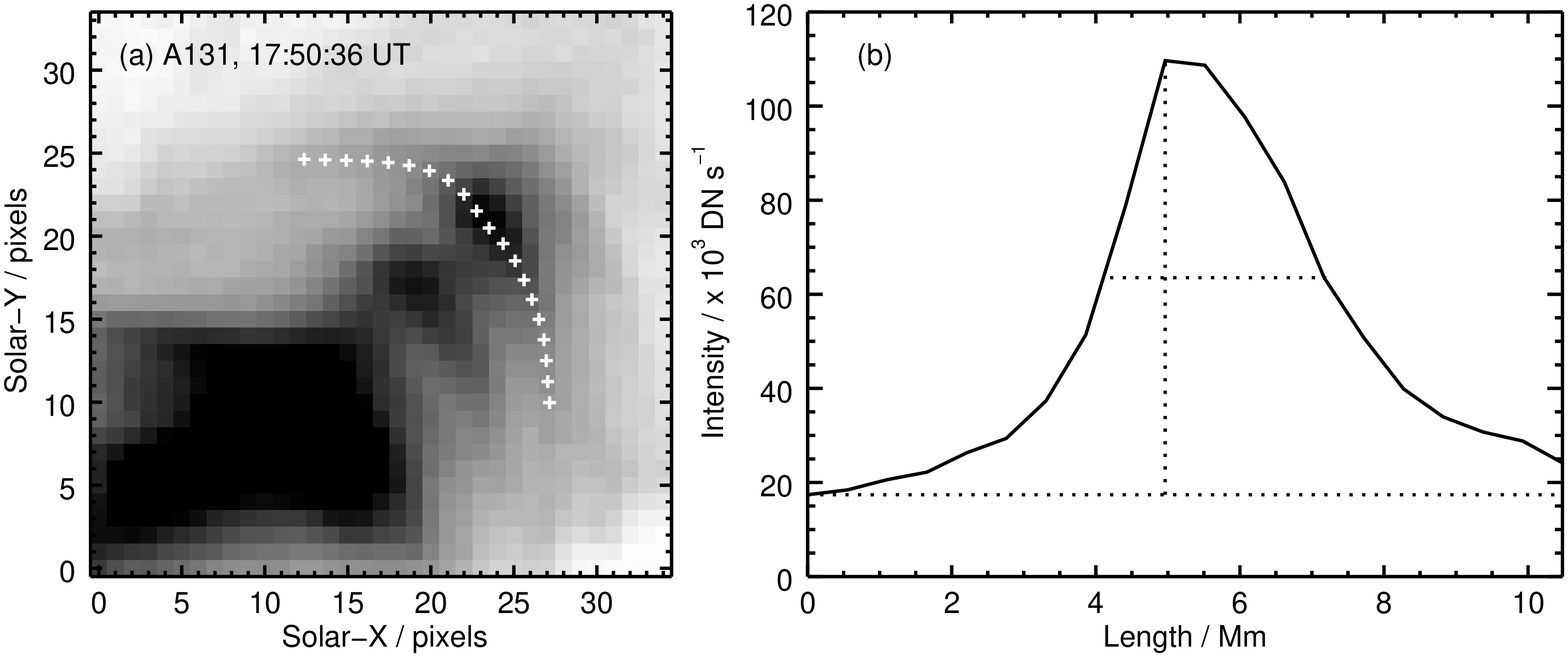}
\caption{Panel (a) shows an A131 image from 17:50:36~UT. A square-root
  intensity scaling has been applied and the color table
  reversed. Interpolated intensities for the spatial locations identified by
  crosses on this image are plotted in Panel (b). Dotted lines show the
  locations of the maximum, minimum and half-maximum intensities.}
\label{fig.arc}
\end{figure}

\ion{Fe}{xxi} emission from the post-flare loop arcade began to dominate at
17:47~UT and Figure~\ref{fig.a131-loops} shows three A131 images
obtained between 17:48 and 17:53~UT.
A striking
feature of these images, particularly at 17:53:00~UT, is that the
loops had bright knots of emission towards the loop apexes. This
feature of post-flare loops has been reported previously from \emph{Yohkoh}
X-Ray Telescope data \citep{acton92,feldman94}.
An asymmetry in the intensity profile across the loop-tops is also
striking, with the
intensity dropping quite sharply on the north sides of the knots, but
more smoothly on the south sides. This is illustrated in
Figure~\ref{fig.arc} where the intensity profile along an arc chosen
to align with a flare loop from the 17:50:36~UT exposure is
plotted. The asymmetry in the intensity distribution is clearly seen,
with the intensity falling to its half-maximum value within 1~Mm on
the north side of the apex, and within 2.2~Mm on the south
side. Intensities along the displayed arc were derived through
bilinear interpolation of the original image. We note that the
exposure time for the 17:50:36~UT is not correct in the AIA data file,
and so the commanded exposure time is used instead. This introduces an
uncertainty of $\pm 7$\%\ in the stated intensities (P.~Boerner, 2014,
private communication).  IRIS spectra allow the properties of the
loop-top brightenings to be compared with the loop legs.

Inspection of the \lam1354.1 profiles in the post-flare loops shows that
they are very close to Gaussian in shape except for the blend with
\ion{C}{i} \lam1354.29, which is present in most spectra. For this
reason we perform automatic Gaussian fits to the IRIS rasters, and
extract intensity, velocity and line broadening parameters for the
loop locations. At each spatial location we perform a three Gaussian
fit to the  wavelength window containing the \ion{Fe}{xxi} line, with
the additional two
Gaussians used for \ion{C}{i} \lam1354.29 and \ion{O}{i}
\lam1355.60. The EIS\_AUTO\_FIT suite of software \citep{eis_sw16} developed for the
Hinode/EIS mission was modified to perform the fitting. This software
allows parameter limits to be applied to the Gaussians to prevent
spurious fits. For example, \ion{Fe}{xxi} \lam1354.1 was restricted to
full-width at half-maximum (FWHM) values between 0.35 and 1.41, while
the much narrower \ion{O}{i} \lam1355.60 line was restricted to FWHM
values between 0.023 and 0.14~\AA.

\begin{figure}[h]
\epsscale{0.7}
\plotone{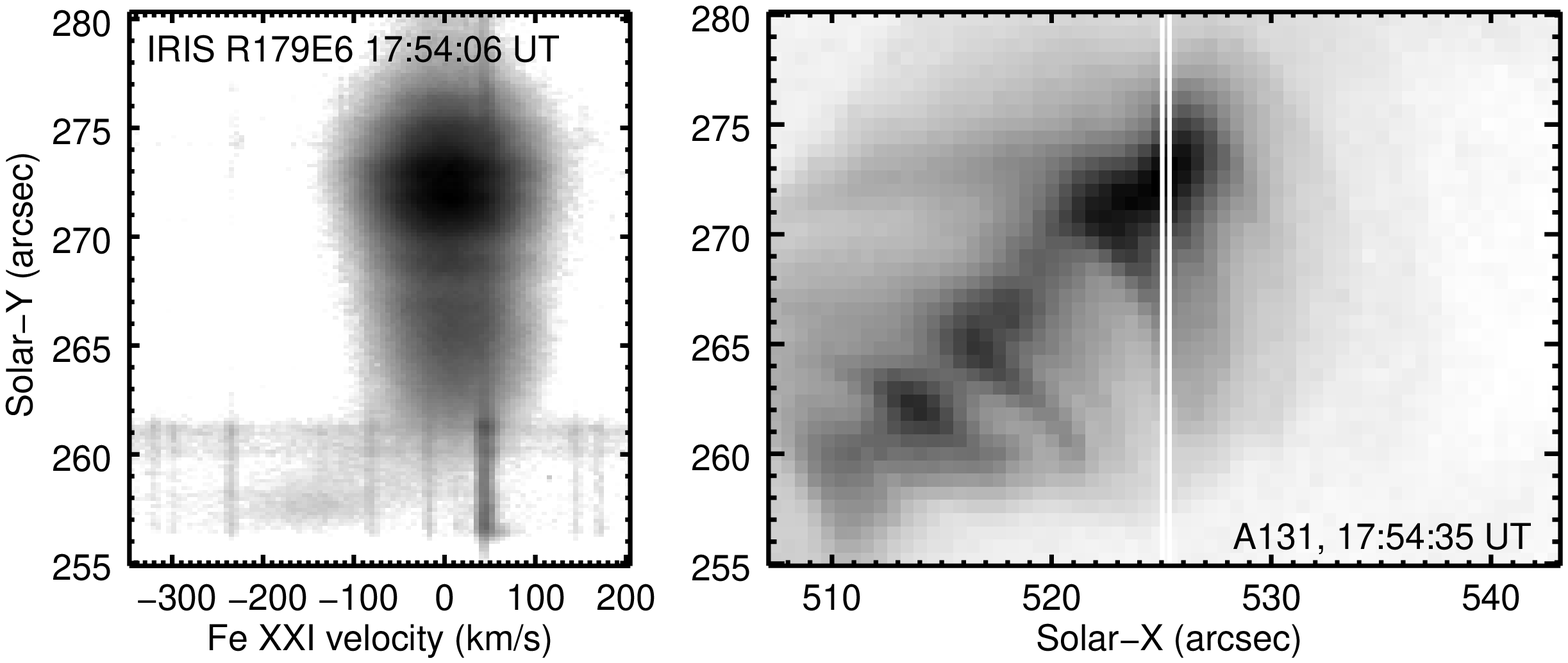}
\plotone{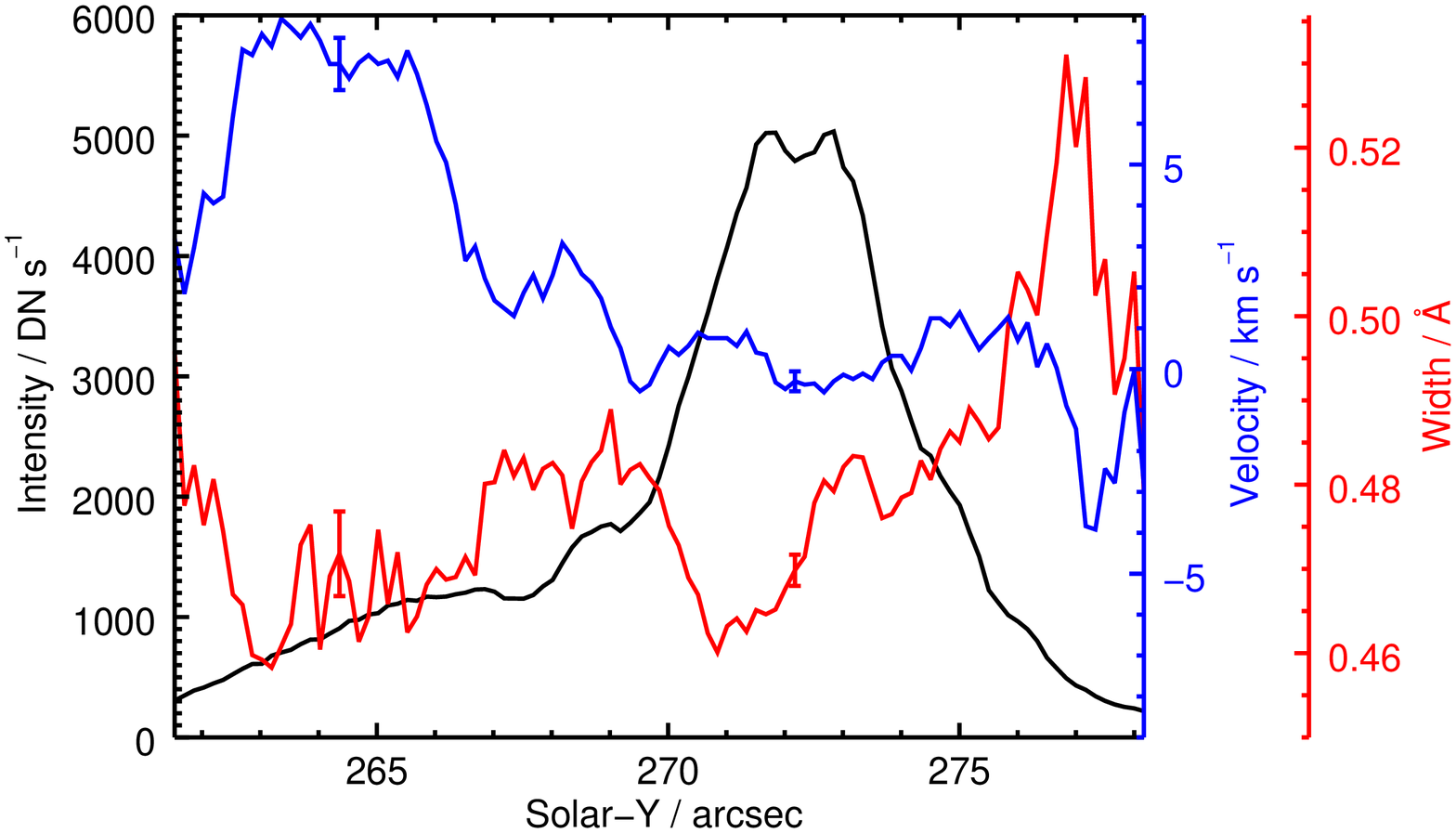}
\caption{\ion{Fe}{xxi} \lam1354.1 Gaussian fit parameters as a function
  of solar-Y position from R179E6 (17:56:06~UT). The black line shows
  the line intensity (DN~s$^{-1}$), the blue line shows LOS velocity
  (\kms), and the red line shows line width (\AA). 
  Fitting uncertainties at example locations are indicated.}
\label{fig.top-leg}
\end{figure}

From the Gaussian fits we can compare how parameters in the bright
loop-tops compare with the loop legs. In Figure~\ref{fig.top-leg} we
show an example from R179E6 (17:54:06~UT). This demonstrates that the
LOS velocity is approximately zero at the loop-top (Y=271--274\as), but
increases to 7--8~\kms\ in the southern loop legs (Y=263--266\as), suggesting a draining of
plasma to the loop footpoints. The line width shows no clear
distinction between the loop-top and the legs, showing that the
loop-top is not hotter than the legs, nor does it have an additional
broadening component that may indicate turbulence. We note that
\citet{jakimiec98} suggested a model for  bright loop-tops whereby
there are tangled magnetic field lines at the loop-top, and the motion
of plasma on these field lines would result in non-thermal broadening
to emission lines. The field is not tangled in the loop-legs in this
model, which would lead to a lower non-thermal broadening in the
emission lines. The IRIS spectra show that the loop-tops do not have
an enhanced broadening compared to the loop-legs.

\begin{figure}[h]
\epsscale{1.0}
\plotone{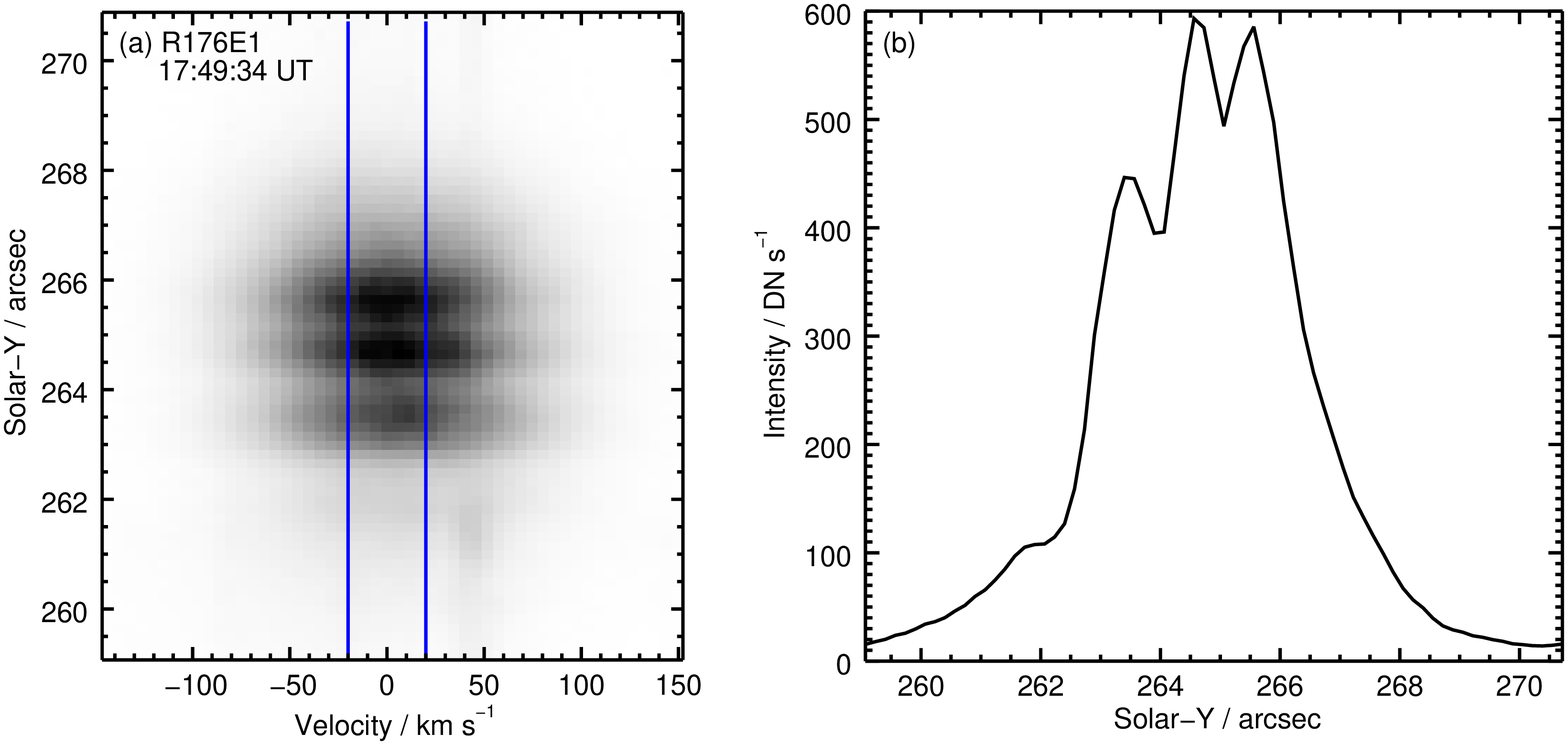}
\caption{Panel (a) shows an IRIS detector image from 17:49:34~UT,
  showing strong \ion{Fe}{xxi} \lam1354.1 emission. Averaging the
  intensity over $\pm$ 20~\kms\ from line center gives the intensity
  profile with solar-Y position shown in panel (b).}
\label{fig.fs}
\end{figure}

IRIS has higher spatial resolution than AIA and so allows the sizes of
the loop-tops to be determined. There are a number of exposures where
quite small features along the IRIS slit can be seen, and an example
from R176E1 (17:49:34~UT) is shown in Figure~\ref{fig.fs}a where
there are three
distinct structures, separated by about 1\as\
(725~km). Averaging the intensity over $\pm 20$~\kms\ either side of
line center gives the intensity profile shown in
Figure~\ref{fig.fs}b, and Gaussian fits to the three structures give
FWHM values of close to 1\as. As this is significantly larger than the
spatial resolution of the instrument, we consider the structures to be
resolved by IRIS. Note that we consider the three structures to be the
loop-tops of three distinct flare loops. Other exposures with similar
examples of fine-scale structure include R176E2, R178E1 and R178E5.

\begin{figure}[h]
\epsscale{0.6}
\plotone{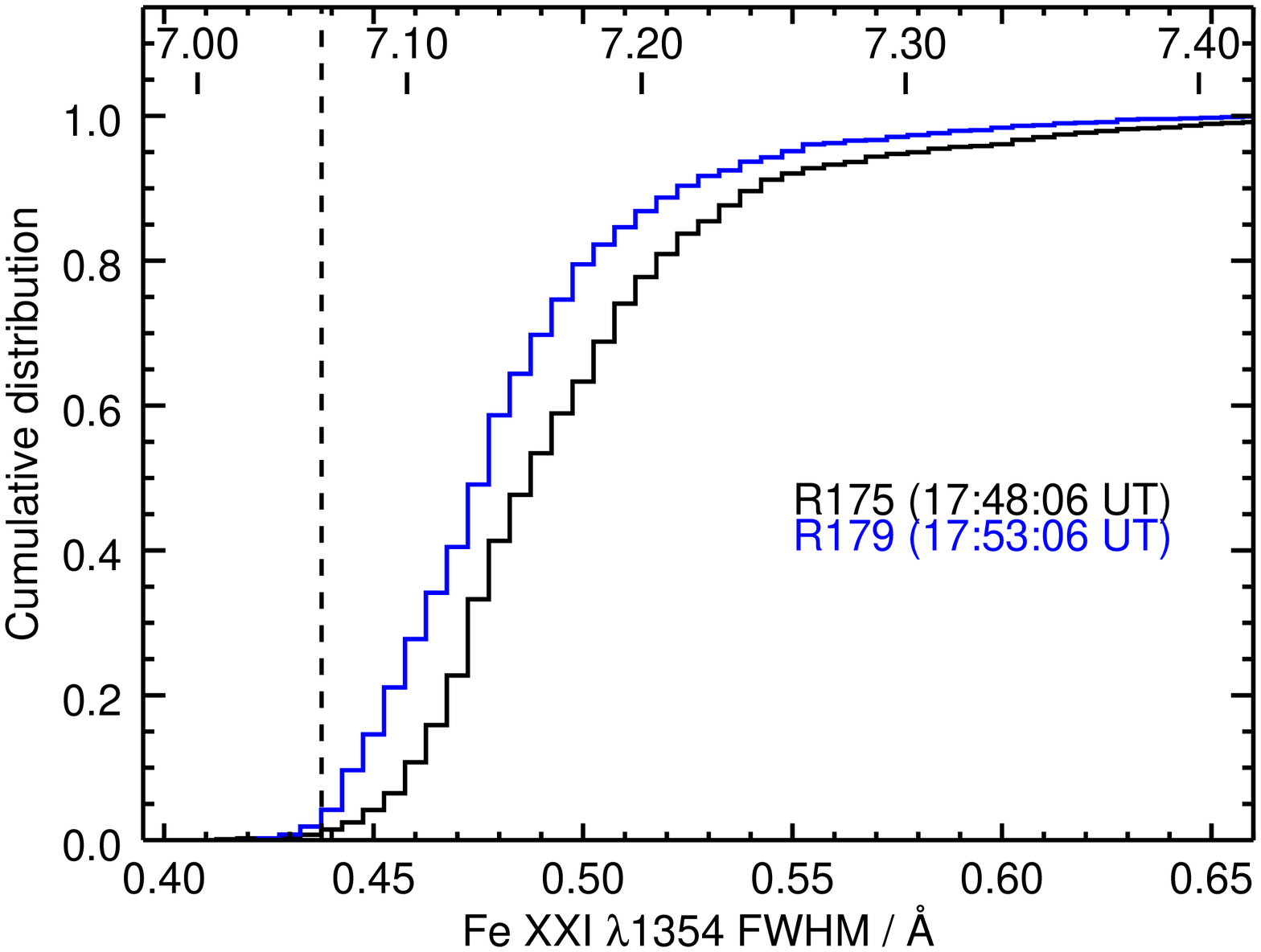}
\caption{Cumulative distributions of \ion{Fe}{xxi} \lam1354.1 line
  widths from rasters R175 and R179. Logarithmic
  temperatures corresponding to the widths, assuming they are entirely
  thermal in origin, are indicated. The vertical dashed line indicates
the thermal width at the peak formation temperature ($\log\,T=7.06$)
of \ion{Fe}{xxi} \lam1354.1.}
\label{fig.wid-hist}
\end{figure}

\citet{doschek75} demonstrated that the \ion{Fe}{xxi} line width
decreased with time for one flare. In the present work we have
spatially-resolved line width measurements and so we take the median
value of the line width, $W$, from each of the rasters R174 to R179, and
these 
are given in Table~\ref{tbl.widths}. The median was applied to between 614 and 1177 spatial
pixels from these rasters. $W$ can be
interpreted as being entirely thermal in origin, in which case a
temperature $T_{\rm th}=1.96\times 10^{12} N\, W/\lambda$~K can be derived,
where $N$ is the mass number of the emitting element. The values of
$T_{\rm th}$ are given in Table~\ref{tbl.widths}. This interpretation
assumes an isothermal plasma, and the results imply cooling as the
flare decays. An alternative interpretation comes from assuming that
\lam1354.1 is formed at its temperature of maximum emission
($\log\,T/{\rm K}=7.06$), in which case line broadening beyond the thermal
width at this temperature corresponds to non-thermal plasma motions.
The non-thermal motions are represented by the velocity $\xi$, given by
\begin{equation}
4 \ln 2 \left( {\lambda\over c} \right)^2 \xi^2 = W^2 - W_{\rm th}^2,
\end{equation}
where $c$ is the speed of light and $W_{\rm th}$ the width at
$\log\,T/{\rm K}=7.06$. Table~\ref{tbl.widths} gives the values of $\xi$ assuming
this interpretation, and they show non-thermal motions decreasing from
43 to 26~\kms\ over a six minute period.
We note that
\citet{doschek75} found peak non-thermal motions of 60~\kms,
falling to 5~\kms\ about 14~minutes later.
In Figure~\ref{fig.wid-hist} we consider the distribution of widths
across two rasters separated by 5~minutes. A cumulative distribution function is used and the
total number of spatial pixels was 830 and 1188 for R175 and R179,
respectively. The distributions demonstrate that, on the whole,
the widths decreased by 0.02~\AA\ over this time, but also that there
are more large widths in the earlier raster: for example, 8.8\%\ of spatial
pixels have widths $\ge 0.55$~\AA\ for R175, compared to 5.3\%\ for
R179.  If the measured widths are interpreted purely as 
thermal widths, then they imply mostly $\log\,T\le 7.35$ for R175 and
$\log\,T \le 7.25$ for R179. We also note there are very few pixels for
which the width is below the thermal width at $\log\,T=7.06$, the
temperature of maximum emission of \lam1354.1, which gives confidence
that this value is accurate.

\begin{deluxetable}{ccccc}
\tablecaption{Average properties derived from \ion{Fe}{xxi} \lam1354.1
  line widths.\label{tbl.widths}}
\tablehead{
Time\tablenotemark{a} & Raster & $W$ & $\xi$ & $\log\,(T_{\rm th}/{\rm K})$ \\
 & & (\AA) & (\kms) &  }
\startdata
17:47:27 & R174 & 0.543 & 42.8 & 7.25 \\
17:48:42 & R175 & 0.493 & 30.0 & 7.16 \\
17:49:57 & R176 & 0.489 & 28.9 & 7.16 \\
17:51:12 & R177 & 0.491 & 29.6 & 7.16 \\
17:52.26 & R178 & 0.482 & 26.9 & 7.14 \\
17:53:41 & R179 & 0.481 & 26.3 & 7.14 \\
\enddata
\tablenotetext{a}{The midpoint time of the raster.}
\end{deluxetable}

\begin{figure}[h]
\epsscale{1.0}
\plotone{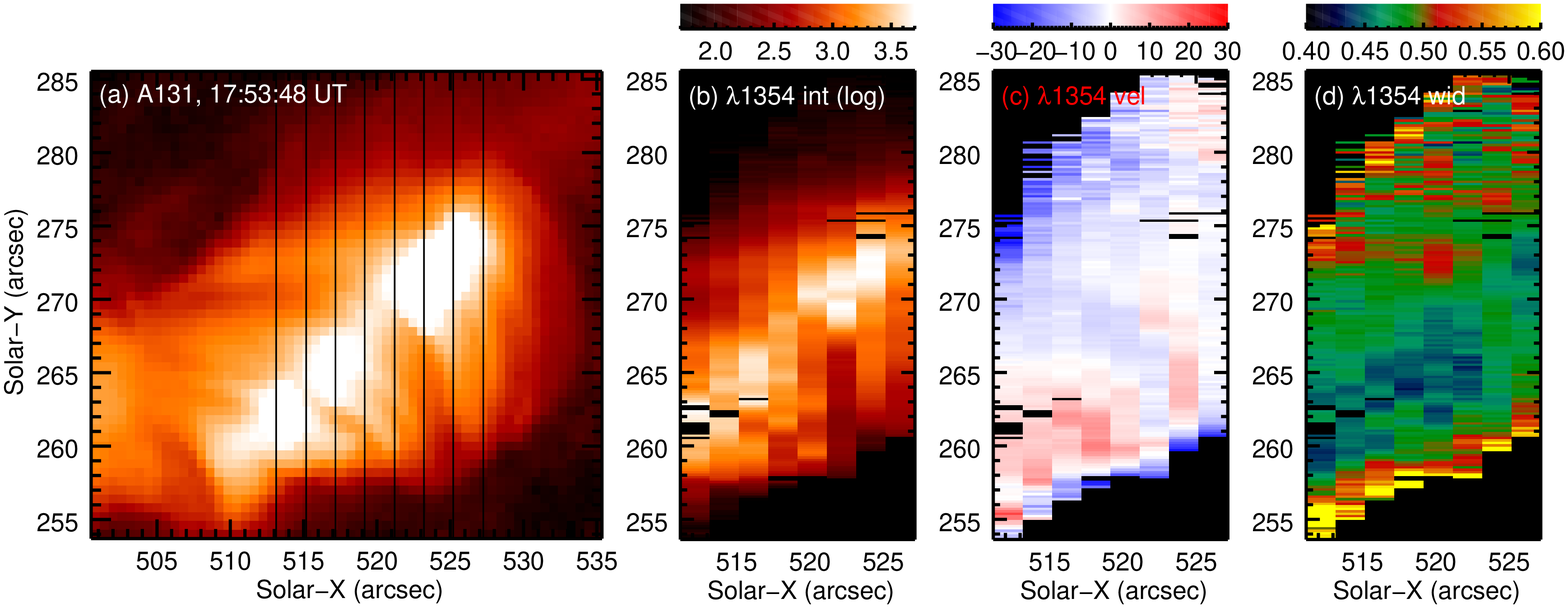}
\caption{Panel (a) shows the A131 image from 17:53:48~UT with a
  logarithmic intensity scaling. The brightest parts of the image are
  saturated, and the eight vertical lines show the
  locations of the IRIS slit for raster R179. Panels (b), (c) and (d)
  show the \ion{Fe}{xxi} \lam1354.1 intensity (logarithm), velocity and width, in
  units of DN~s$^{-1}$, \kms, and \AA, respectively.}
\label{fig.vel}
\end{figure}

Away from the ribbons, the Doppler shifts of \lam1354.1 are modest and
in fact the IRIS data can be used to estimate a new reference
wavelength of 1354.106~\AA\ for the line
(Appendix~\ref{sect.ref}).  Figure~\ref{fig.vel} shows intensity,
velocity and line width maps derived from IRIS raster R179, compared
to a co-temporal A131 image. Note that a pixel size of 2\as\ in the
X-direction is used for displaying the IRIS images but in reality each
column represents only a region 0.33\as\ in the X-direction. The
velocity map is derived assuming the rest wavelength of 1354.106~\AA\
for the \ion{Fe}{xxi} line (Appendix~\ref{sect.ref}). If the value of
1354.064~\AA\ from \citet{feldman00} is used, then all velocities
would be increased by 9~\kms, making the loop arcade almost entirely
red-shifted. Figure~\ref{fig.vel}c shows that the region around
spatial location (516,260)
 is red-shifted by around 10--20~\kms\ and this pattern
persists throughout rasters R174 to R179. The northern loop legs
generally show a weak blue-shift of a few \kms, but with an
uncertainty of $\pm 5$~\kms\ in the rest wavelength of the
\ion{Fe}{xxi} line (Appendix~\ref{sect.ref}) this is not significant.

\section{Conclusions}\label{sect.summary}

In this work we have presented observations of the \ion{Fe}{xxi}
\lam1354.1 emission line obtained by IRIS during the 2014 March 29 X1
flare. The high spatial resolution and sensitivity of the IRIS
instrument allows high temperature ($\approx 10$~MK) plasma to be
studied on much smaller spatial scales than previously possible. This
has enabled fine structure in the post-flare loop arcade and the flare
ribbons to be studied. 
We distinguish the \ion{Fe}{xxi} emission at the flare ribbons from
that in the post-flare loop arcade, and we consider that the ribbons
represent the footpoints of the post-flare loops. 

The variety of \ion{Fe}{xxi} profiles at and near the ribbon sites is
complex. There are four distinct ribbons, labeled N1, N2, S1 and S2
(see Figure~\ref{fig.2796}), with the
brighter N1 and S2 ribbons behaving in the classical manner, being
roughly parallel to the polarity inversion line and moving apart with
time. These two ribbons both show blue-shifted \ion{Fe}{xxi} emission
at or close-to the chromospheric ribbon sites, in agreement with the standard flare model. 
The
N2 ribbon sites are partly or wholly compromised by overlying flare loop
plasma. The S1 ribbon is unusual as it is of the same polarity as S2
and roughly parallel to it, but it does not move outwards away from
the polarity inversion line as S2 does. It also does not show
blue-shifted \ion{Fe}{xxi} emission, although observations are
affected by overlying loop plasma. If we treat the N2 and S1 ribbons
as anomalous, then the results for the N1 and S2 ribbons can be
summarized as:
\begin{enumerate}
\item Of the nine raster positions on the N1 and S2 ribbons, all of
  them show \ion{Fe}{xxi} emission at or close-to the ribbon sites. Eight
  show \ion{Fe}{xxi} blue-shifts of 100~\kms\ or more.
\item The \ion{Fe}{xxi} emission appears at the same time as the
  ribbon emission for three of the nine positions. There is a delay of
  75~seconds in four cases, and a delay of 150~seconds in one
  case. (The time resolution is restricted by the raster cadence of
  75~seconds.)
\item The \lam1354.1 emission when it first appears at the ribbons is mostly compact with a
  spatial extent $<2$\as.
\item The S2 ribbon moves southwards by up to 10\as\ during the IRIS
  raster. \ion{Fe}{xxi} emission is found at the previous locations of
  the ribbon, but the intensity is patchy and the velocity
  varies. There is some ambiguity as to whether the emission extends
  along the leg of a single loop, or whether it is low-lying footpoint
  emission from multiple loops.
\item At some of the ribbon locations, blue-shifted \ion{Fe}{xxi}
  emission remains present until the end of the raster sequence,
  although the speeds decrease to around 20--60~\kms.
\item Studies of \lam1354.1 at the ribbons are compromised by a number
  of chromospheric emission lines that are found between 1352.5 and
  1354.0~\AA\ at the ribbon locations and are comparable in strength
  to \lam1354.1.
\end{enumerate}

The post-flare loop \lam1354.1 emission began to appear during the impulsive
phase of the flare and developed into an extended loop arcade with
bright loop-tops. The key results are listed below.
\begin{enumerate}
\item The AIA 131~\AA\ filter images, which are dominated by \ion{Fe}{xxi} \lam128.7,
  show that the bright loop-tops have an asymmetric intensity
  distribution, being more extended on the south-side of the loops.
\item The loop-tops are resolved by IRIS and have sizes of $\ge
  1$\as. 
\item IRIS Doppler maps formed from \lam1354.1 show small Doppler
  shifts in the loop arcade, corresponding to velocities of typically
  $\le 10$~\kms.
\item Velocities at the loop-tops are close to rest, and the \lam1354.1
  width is not significantly enhanced compared to the loop legs.
\item On average there is a slow decrease in the width of the \lam1354.1
  emission line of about 0.02~\AA\ over 5~minutes, and there is a
  greater dispersion of widths near the peak of the flare.
\item Assuming the loop arcade emission is, on average, at rest then a
  new rest wavelength of $1354.106\pm 0.023$~\AA\ is derived for the
  \ion{Fe}{xxi} line.
\end{enumerate}

The observations generally show agreement with features of the
standard solar flare model. In particular \ion{Fe}{xxi} is found at
the ribbon sites as expected from the intense heating occurring there,
and the line is mostly blue-shifted as expected from models of
chromospheric evaporation \citep[e.g.,][]{fisher85a}. The connection
between the footpoints and  the post-flare loops is less clear partly
because of the sparse spatial sampling of the rasters. There are a
number of exposures such as R179E6 (Figure~\ref{fig.top-leg}) where
blueshifts at the ribbon are observed simultaneously with bright,
stationary loop plasma, but a smooth transition from blue-shift to
stationary velocity is not common, the intensity often being somewhat
patchy at both the footpoint regin and along the loops. This may
result from the slit being inclined to the loops' axes causing
multiple loops to be observed in one exposure.

The IRIS instrument has significant flexibility in terms of
observation programs and so we highlight here some of the advantages and
disadvantages of the observing sequence used for the March~29
flare. Firstly, the eight second exposure time is ideal for studying
\ion{Fe}{xxi} \lam1354.1 as it yields sufficient signal to study the
weak profiles seen at the ribbons, but also the intensities in the
post-flare loops do not reach sufficiently high levels  that saturation sets
in. In particular we note that if the exposure times had been reduced
by automatic exposure control then this would have made studying the
ribbon emission difficult.

The raster type used for the observation was a ``coarse 8-step
raster'', which had 2\as\ jumps between slit positions. This enabled a
fairly large spatial region to be covered rapidly in X, with the
downside of poor sampling. As the IRIS slit did not lie directly along
the axes of the post-flare loops, the jumps in X-position generally
meant that only small sections of the loops could be observed
spectroscopically. In particular the footpoint emission shown in
Figure~\ref{fig.closeups} generally terminated abruptly in Y a short
distance from the ribbon. A ``dense raster'', i.e., step sizes equal
to the slit width would have enabled us to check if the variation of footpoint
emission with height along the loops could be tracked in the
X-direction. 

Finally, the wavelength window used for \lam1354.1 is not sufficiently
large, with at least one profile (Figure~\ref{fig.profiles}c) partially
extending beyond the window edge at $-350$~\kms. Presently there are
five options offered to observers for line-lists, and the ``flare
line-list'' was used for this observation. Extending the 1354~\AA\
window to at least $-500$~\kms\ from line center is recommended. Also
extending the window by around $50$~\kms\ on the long wavelength side
would also be beneficial as \ion{O}{i} \lam1355.60 (a line needed for
wavelength calibration) is partly out of the window at some locations.

\acknowledgments

P.R.Y.\ acknowledges funding from NASA grant NNX13AE06G and
National Science Foundation grant 
AGS-1159353. H.T.\ is supported by contracts 8100002705 and SP02H1701R
from LMSAL to SAO.
IRIS is a NASA small explorer mission developed and operated by LMSAL
with mission operations executed at NASA Ames Research center and
major contributions to downlink communications funded by the Norwegian
Space Center (NSC, Norway) through an ESA PRODEX contract. SDO is a
mission for NASA's Living With a Star 
program, and data are provided courtesy of NASA/SDO and the AIA and
HMI science teams.

{\it Facilities:} \facility{IRIS}, 
\facility{SDO(AIA)}, \facility{GOES}.

\appendix

\section{Blending lines near Fe XXI \lam1354.1}\label{sect.blend}

\begin{figure}[h]
\epsscale{0.6}
\plotone{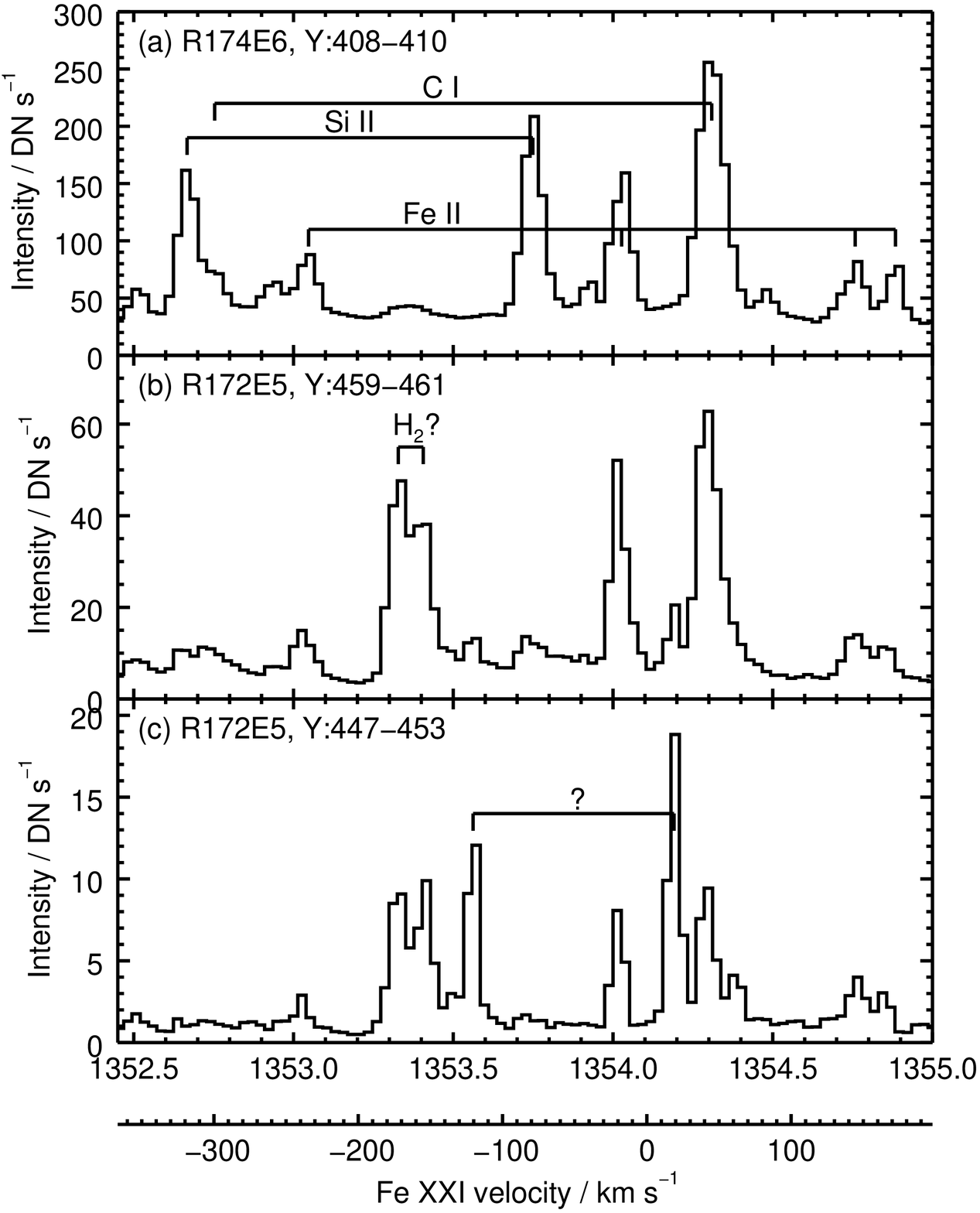}
\caption{Three examples of IRIS spectra at or in the vicinity of flare
ribbons. The Y-pixel ranges over which the spectra are averaged are indicated. }
\label{fig.blend2}
\end{figure}

In flare spectra, particularly near the ribbons, a number of emission
lines become prominent that are not normally measurable in the IRIS
spectra. Some of these lines compromise measurements of \ion{Fe}{xxi}
\lam1354.1 and so we discuss these lines here. We note that the large
width of \lam1354.1 compared to the cool blending lines means that it
is relatively easy to identify \lam1354.1, but the lines affect the
measurement of the line parameters (intensity, centroid and width). As discussed in
Sect.~\ref{sect.fe21r}, the ribbons typically appear in chromospheric
lines one raster earlier than they do in \lam1354.1, so in
Figure~\ref{fig.blend2} we show  examples of ribbon spectra without
the \lam1354.1 line in order to better display the blending
lines. These spectra can be compared with Figure~\ref{fig.profiles}
which shows four examples where 
\lam1354.1 is present.


The spectrum in Figure~\ref{fig.blend2}a is the most common type of ribbon
spectrum, with two lines of \ion{Si}{ii} and one of \ion{Fe}{ii}
becoming comparable in intensity to \ion{C}{i} \lam1354.29.
The two \ion{Si}{ii} lines often show interesting dynamics at the
flare ribbons, with broad long-wavelength wings extending up to
150~\kms. The broad wings can be confused with \ion{Fe}{xxi} emission,
but the presence of two \ion{Si}{ii} lines enables the \ion{Si}{ii}
components to be easily identified.

Figure~\ref{fig.blend2}b shows a spectrum about 1\as\ from a ribbon
where a pair of very close lines are found at 1353.32 and
1353.39~\AA. By comparing the intensity distribution along the slit,
these lines show similar behavior to a pair of lines at 1333.48 and
1333.80~\AA\ that are known to be H$_2$ lines fluoresced by
\ion{Si}{iv} \citep{bartoe79}. Another possibility is that they are CO
lines, although they are not listed by \citet{jordan79}. 

An unusual spectrum is seen in a few of the flare exposures and an
example is shown in Figure~\ref{fig.blend2}c. Two lines at 1353.55
and 1354.19\AA\ are particularly prominent and are striking because
of their very narrow widths of $\approx 0.045$~\AA. The lines have a
very different intensity distribution along  the slit compared to the
atomic lines or H$_2$ and there are many similar lines in the spectrum
including a distinctive group of 5--6 lines between 1402.0 and
1402.5~\AA, on the short wavelength side of \ion{Si}{iv} \lam1402.77.

\section{The rest wavelength of Fe\,XXI \lam1354.1}\label{sect.ref}

For creating velocity maps from \ion{Fe}{xxi} \lam1354.1 it is necessary
to set an absolute wavelength scale. For this we use \ion{O}{i}
\lam1355.60 which is a narrow chromospheric line that demonstrates
only small Doppler shifts in rasters. The reference wavelength for this
line is 1355.5977~\AA, which is a calculated value with an accuracy
estimated at 0.5~m\AA\ \citep{eriksson68}. As described in Sect.~\ref{sect.fe21l} we fit three
Gaussians to the wavelength window containing \ion{Fe}{xxi} \lam1354.1,
one of which represents the \ion{O}{i} line. 

We used rasters R177--179 to determine the \ion{Fe}{xxi}
wavelength as these show strong emission from the post-flare loops,
and the line is mostly close to the rest wavelength. We restrict
analysis to those spatial pixels for which the \lam1354.1 line width is
between 0.42 and 0.55~\AA, i.e., close to the thermal width of
0.43~\AA, the integrated intensity is $\ge$ 50~DN, and the reduced
$\chi^2$ value for the fit is $\le$ 1.5. These restrictions remove
pixels for which the line shows unusual dynamics,  is weak, or has a
poor fit. 
For these spatial pixels we take the set of \ion{O}{i} centroids,
measured through a Gaussian fit, and compute the mean and standard
deviations. These values are shown in Table~\ref{tbl.ref}. Similarly
the mean and standard deviation of the \ion{Fe}{xxi} wavelengths are
computed. We correct the derived \ion{Fe}{xxi} wavelength by the
difference between  the derived and reference \ion{O}{i} wavelengths,
giving the rest \ion{Fe}{xxi} wavelengths shown in
Table~\ref{tbl.ref}. The uncertainty has been obtained by adding in
quadrature the standard deviations of the measured \ion{Fe}{xxi} and \ion{O}{i}
wavelengths, and the uncertainity in the \ion{O}{i} reference
wavelength.  Combining these results gives a final rest wavelength of
$1354.106\pm 0.023$~\AA, which we adopt in the present work.

We note that there is a significant discrepancy with the value of $1354.064\pm
0.020$~\AA\ from \citet{feldman00}. If this value is correct, then it
implies that the post-flare loops in the present observation are, on
average, red-shifted by 9~\kms.

\begin{deluxetable}{lcc}
\tablecaption{Measured wavelengths for \ion{O}{i} and \ion{Fe}{xxi}.\label{tbl.ref}}
\tablehead{
  Raster & \ion{O}{i}  & \ion{Fe}{xxi}
}
\startdata
R177 & $1355.6204 \pm 0.0090$ & $1354.106 \pm 0.023$ \\
R178 & $1355.6205 \pm 0.0080$ & $1354.106 \pm 0.022$ \\
R179 & $1355.6211 \pm 0.0078$  & $1354.107 \pm 0.023$ \\
\enddata
\end{deluxetable}

\bibliographystyle{apj}
\bibliography{myrefs}

\end{document}